\documentclass[amsmath,amssymb,aps,prd,showkeys,reprint, floatfix,groupedaddress,superscriptaddress,amsmath,amssymb]{revtex4-1}

\usepackage[breaklinks = true,colorlinks = true,linkcolor = blue,citecolor = blue]{hyperref}
\usepackage{graphicx}

\begin{document}

\title{Fluctuations and phases in baryonic matter}

\author{Len Brandes}
\email{len.brandes@hotmail.de}
\affiliation{Physics Department, Technical University of Munich, 85748 Garching, Germany}

\author{Norbert Kaiser}
\email{nkaiser@ph.tum.de}
\affiliation{Physics Department, Technical University of Munich, 85748 Garching, Germany}

\author{Wolfram Weise}
\email{weise@tum.de}
\affiliation{Physics Department, Technical University of Munich, 85748 Garching, Germany}

\begin{abstract}
The phase structure of baryonic matter is investigated with focus on the role of fluctuations beyond the mean-field approximation. The prototype test case studied is the chiral nucleon-meson model, with added comments on the chiral quark-meson model.  Applications to the liquid-gas phase transition in nuclear matter and extensions to dense matter are performed. The role of vacuum fluctuations and thermal excitations is systematically explored. It is pointed out that such fluctuations tend to stabilize the hadronic phase characterized by spontaneously broken chiral symmetry, shifting the chiral restoration transition to very high densities. This stabilization effect is shown to be further enhanced by additional dynamical fluctuations  treated with functional renormalisation group methods. 
\end{abstract}

\maketitle

\section{Introduction}
\label{sec:intro}

The QCD phase diagram in the region of high baryon densities and low temperatures is still one of the great unknowns in the physics of the strong interaction. Its behaviour at high temperature and small baryon chemical potentials is quite well understood from lattice QCD thermodynamics \cite{Bazavov2014} and from the analysis of high-energy heavy-ion collisions (for recent reviews see \cite{PBM2016,Andronic2018} and references therein). It is interpreted as a continuous crossover from the hadronic to the quark-gluon phase around a transition temperature $T_c\simeq 155$ MeV. On the other hand, extensions of the phase diagram to high densities at low temperatures from first-principles theory are hindered by the notorious sign problem of lattice QCD \cite{Muroya2003,Aarts2016}. A key question in this context concerns the possible existence of a first-order phase transition from spontaneously broken to restored chiral symmetry at high density.

Empirically, the existence of heavy neutron stars \cite{Oezel2016} with masses around and even above 2$M_\odot$ \cite{Demorest2010,Antoniadis2013,Fonseca2016,Cromartie2019} sets strong constraints on the equation-of-state (EoS) of dense baryonic matter \cite{Oertel2017}. This EoS must be sufficiently stiff, i.e. the pressure $P(\cal{E})$ at energy densities $\cal{E}\sim$ 1 GeV/fm$^3$ must be large enough to support such massive compact objects against gravitational collapse. The detection of gravitational wave signals from two merging neutron stars \cite{Abbott2017}  adds further important information on the EoS, by providing limits for the tidal deformability and for neutron star radii \cite{Most2018,De2018}. For recent analyses constraining the dense matter EoS using data from LIGO/Virgo together with NICER measurements, see Refs. \cite{Raaijmakers2020,Essick2020}.

A broad variety of models has been developed over decades to address the EoS of dense baryonic matter. A principal requirement for any such model to be acceptable is its capability of reproducing properties of nuclear matter consistent with empirical phenomenology around the nuclear equilibrium density, $n_0 = 0.16$ fm$^{-3}$. An early successful example of this kind is the variational APR model \cite{APR1998}. Its basic degrees of freedom are nucleons interacting through pion exchanges plus phenomenological short-distance two- and three-body forces. In this approach repulsive correlations, with their continuously rising strength as the baryon density increases, produce a sufficiently stiff EoS that is able to support even the most massive observed neutron stars. 

More recent theoretical developments are guided by the approximate chiral symmetry of QCD in its two-flavour ($u$ and $d$ quark) sector. Spontaneously broken chiral symmetry at low energies implies that a chiral effective field theory of pions as Nambu-Goldstone bosons, coupled to nucleons as ``heavy" fermions, is a valid framework for treating the nuclear many-body problem and its thermodynamics at sufficiently low densities and temperatures \cite{Holt2013,Krueger2013,Wellenhofer2015,Drischler2016,Holt2016,Drischler2021}. Such perturbative approaches give reliable descriptions of both nuclear and neutron matter up to about twice the density of normal nuclear matter, $n\lesssim 2\, n_0$. At higher densities non-perturbative methods are required. This is where chiral nucleon-meson \cite{Floerchinger2012,Drews2014,Drews2015,Drews2017} or quark-meson \cite{Schaefer2005,Schaefer2007,Herbst2011,Chatterjee2012,Gupta2012,Tripolt2018,Zacchi2018,Pereira2020} model Lagrangians have frequently been used to start with, treated in mean-field approximation or in combination with functional renormalisation group (FRG) methods. 

Matter in the core of neutron stars has always been in the focus of computations and extrapolations into the region of the highest observable baryon densities. Given the present empirical constraints, it is generally believed that densities 
of order $5-6\,n_0$ are reached in the neutron star central regions \cite{Hell2014,Jiang2019}. Such densities are not ultra-high, in the sense that the corresponding mean distance between baryons (mostly neutrons with a fraction of protons) is still about 1 fm, so that an interpretation of the EoS in terms of nucleon quasiparticles within relativistic Landau Fermi liquid theory \cite{FW2019} is still meaningful. But at the same time this is the range of densities at which the quark cores of the nucleons begin to touch and overlap. Hybrid scenarios characterised by a continuous crossover from baryon to quark degrees of freedom have been designed in this context \cite{Baym2018,Baym2019,McLerran2019,Jeong2020,Fujimoto2020a,Fukushima2020}. At even higher densities various forms of colour superconductivity are expected to take over until, at asymptotically large Fermi momenta, perturbative QCD can be applied and sets limiting conditions for extrapolations of the EoS \cite{Annala2018,Vuorinen2019}. Resummed QCD perturbation theory \cite{Fujimoto2020b} permits lowering the baryon density from extreme limits and favors a smooth matching to the EoS at typical neutron star central densities.

In all these and related considerations, a possible first-order chiral phase transition, and the quest for a corresponding critical end point in the QCD phase diagram, have always been themes of prime interest \cite{Stephanov1998,Fukushima2003,Stephanov2004,Fischer2014}. As mentioned, lattice QCD with its limitation to small chemical potentials cannot fundamentally clarify these issues, and there is so far no empirical evidence for a critical end point. Earlier hypotheses for the existence of such a first-order phase transition were primarily based on mean-field (MF) calculations using Nambu \& Jona-Lasinio (NJL) type models \cite{Asakawa1989,Klimt1990,Scavenius2001}, often extended by adding some confinement aspects through the Polyakov loop (PNJL) \cite{Roessner2007,Roessner2008,Hell2010}. NJL and PNJL model studies \cite{Kitazawa2002,Fukushima2008,Abuki2010,Bratovic2013} already indicated that the existence and properties of the chiral phase transition are highly sensitive to variations in the strengths of vector couplings and of the axial $U(1)$ breaking interaction. Furthermore, results of alternative chiral models \cite{Drews2015,Drews2017,Chatterjee2012,Gupta2012,Tripolt2018,Zacchi2018} pointed out that the chiral phase transition and thermodynamics at low temperature are strongly influenced by the treatment of fluctuations beyond MF. 

It is this latter point that we wish to investigate in more detail in the present work: how does the nature of a possible chiral phase transition in dense baryonic matter depend on effects of fluctuations beyond mean-field approximation? For demonstration we shall use mostly the chiral nucleon-meson (ChNM) model. Some comments will also be added concerning the chiral quark-meson (ChQM) model. Each of these models has its merits and limitations. The ChNM model is able to describe nuclear and neutron-rich matter realistically within the phase of spontaneously broken chiral symmetry, typically up to a few times the density of normal nuclear matter. The empirical signatures of nuclear thermodynamics including the liquid-gas phase transition establish the input for fixing parameters of the ChNM model that is then extended to higher densities. The ChQM model is more schematic and restricted in its applicability as it misses the localisation and clustering of quarks into nucleons. Hence, it cannot be used in density regions where nucleons and their interactions dominate. But it has been useful in providing some guidance and insights at higher densities when approaching the chiral restoration transition.  

This paper is organized as follows. Section II gives a brief account of vacuum and other fluctuations and their treatment using  functional renormalisation group methods. Section III discusses chiral order parameters in nuclear and neutron matter, their mean-field characteristics and their behaviour in the presence of fluctuations beyond MF, using primarily the ChNM model  as a prototype representative of chiral effective theories with baryons. Section IV extends the analysis to a chiral quark-meson model. A summary and conclusions are presented in Section V. 

\section{Mean field and beyond in chiral models}

This section briefly introduces some basics of the field theoretical model and the schemes to be examined: mean-field (MF) approximation, and fluctuations beyond MF as they emerge from a functional renormalisation group (FRG) treatment. Fermionic vacuum fluctuations will be given separate attention following Ref.\,\cite{Skokov2010} where it was demonstrated how they can be treated elegantly within an extended renormalized mean-field (EMF) framework. It will be shown that these vacuum fluctuations already shift the chiral phase transition pattern significantly. FRG calculations naturally include vacuum fluctuations as part of a larger class of fermionic and bosonic loop effects which further modify the chiral restoration scenario at high baryon density.

As a basic framework we consider a chiral theory of fermion doublets (here: nucleons, $\Psi = (p,n)$)
coupled to a chiral boson field $\phi = (\sigma,\boldsymbol{\pi})$ composed of a heavy scalar $\sigma$, and the pion $\boldsymbol{\pi}$ as the pseudoscalar Nambu-Goldstone boson of spontaneously broken chiral $SU(2)_L\times SU(2)_R$ symmetry. In addition to examining fermionic vacuum loop effects, the aim is to study the role of fluctuations of the chiral (pion and scalar) fields in the presence of the filled Fermi sea of nucleons, as one moves to high baryon densities. 

\subsection{Chiral nucleon-meson model}

We start with the (Euclidean) Lagrangian of the ChNM model \cite{footnote0}:
\begin{eqnarray}
{\cal L} &=&\bar{\Psi}\left[\gamma_\mu\partial_\mu + g(\sigma + i\gamma_5\,\boldsymbol{\tau\cdot\pi})\right]\Psi \nonumber \\
&+& {\frac12}\left(\partial_\mu \sigma \partial_\mu \sigma + \partial_\mu \boldsymbol{\pi}\cdot\partial_\mu \boldsymbol{\pi}\right) + {\cal U}(\sigma, \boldsymbol{\pi}) + \Delta{\cal L}~.
\label{eq:Lagrangian}
\end{eqnarray}
The potential ${\cal U}(\sigma, \boldsymbol{\pi})$ is written as a polynomial in the chiral invariant, $\chi \equiv{1\over 2}\phi^\dagger\phi = {1\over 2}\left(\sigma^2 + \boldsymbol{\pi}^2\right)$, and a symmetry breaking piece proportional to the squared pion mass, $m_\pi^2$: 
\begin{equation}
{\cal U}(\sigma, \boldsymbol{\pi}) = \sum_{n = 1}^{N}{a_n\over n!}\left(\chi-\chi_0\right)^n 
- m_\pi^2 f_\pi\left(\sigma-\langle\sigma\rangle_{vac}\right)~.
\label{eq:potential}
\end{equation}
The default maximum polynomial order is chosen as $N = 4$.
Here $\chi_0 = {1\over 2}\langle\sigma\rangle_{vac}^2$ is the vacuum expectation value of the chiral $\chi$ field, with  $\langle\boldsymbol{\pi}\rangle = 0$ in the assumed absence of a pion condensate.  The normalisation of the vacuum scalar field is $\langle\sigma\rangle_{vac} = f_\pi\simeq 93$ MeV, the pion decay constant in vacuum \cite{footnote1}.

The $\Delta{\cal L}$ part of the Lagrangian (\ref{eq:Lagrangian}) is introduced to deal with short-distance dynamics, expressed in terms of isoscalar and isovector vector fields, $v_\mu$ and $\boldsymbol{w}_\mu$, coupled to nucleons. These massive vector fields are treated as homogeneous, time-independent background fields:
\begin{equation}
\Delta{\cal L} = \bar{\Psi}\left[-i\gamma_\mu (g_v v_\mu + g_w\,\boldsymbol{\tau\cdot w_\mu})\right]\Psi  +{1\over 2} m_v^2\left(v_\mu^2 + \boldsymbol{w}_\mu^2\right)~.
\end{equation}
A common mass scale $m_v$ of order 1 GeV is assigned to both isoscalar and isovector boson fields. Note that these auxiliary vector fields need not necessarily be identified with the physical $\omega$ and $\boldsymbol{\rho}$ mesons. Their large mass implies that, for momentum scales with $q^2 < m_v^2$, the resulting short-range Yukawa interactions between nucleons can be treated in mean-field approximation, neglecting fluctuations. In practice, for an isotropic medium, only rotationally invariant solutions of the (static) vector field equations matter and the space components vanish, $v_i =0$ and $\boldsymbol{w}_i = 0$ for $i = 1,2,3$. It turns out to be convenient to rewrite the remaining Euclidean time components $v_4$ and $\boldsymbol{w}_4$ in terms of (real) Minkowskian zero-components, $v_4 = -i v^0$ and $\boldsymbol{w}_4 = -i\boldsymbol{w}^0$. Finally, the mean-field treatment of the vector fields implies that only the isospin-3 component, $w_3^0$, of the isovector field $\boldsymbol{w}^0$ matters. We simply denote those remaining fields as $v^0 \equiv v$ and  $w_3^0\equiv w$ in the following. Then we have
\begin{equation}
\Delta{\cal L} = -\Psi^\dagger \left[g_v\,v + g_w\,\tau_3 \,w\right]\Psi - {1\over 2} m_v^2\left(v^2 + w^2\right)~.
\label{eq:vector}
\end{equation}

The potential ${\cal U}(\sigma, \boldsymbol{\pi})$ and the short-distance terms $\Delta{\cal L}$ are constructed such as to be consistent with selected ground state properties of nuclear matter. For more details see Refs.\,\cite{Floerchinger2012,Drews2015,Drews2017}. In the present work we give an updated and improved parametrisation of these terms.

The Yukawa coupling $g$ in Eq.(\ref{eq:Lagrangian}) is fixed by the nucleon mass $M_N$ in vacuum through the relation
\begin{equation}
M_N = g\langle\sigma\rangle_{vac} = g\,f_\pi~.
\label{eq:Nmass}
\end{equation} 
With $M_N=0.939$ GeV and $f_\pi = 93$ MeV, we have $g\simeq 10.1$. 

\subsubsection{Mean-field thermodynamics}

In the MF approximation the chiral fields $\sigma$ and $\boldsymbol{\pi}$ are replaced by their expectation values,
$\langle\sigma\rangle$ and $\langle\boldsymbol{\pi}\rangle$ (with $\langle\boldsymbol{\pi}\rangle =0$, again assuming the absence of a pion condensate). The MF partition function $Z_{MF}$, or equivalently, the grand canonical potential $\Omega_{MF}$ as function of temperature $T$ and chemical potentials $\mu_{p,n}$ of proton and neutron, becomes:
\begin{eqnarray}
\Omega_{MF} &=& - {T\over V}\ln Z_{MF} = \Omega_F(T,\mu_p,\mu_n;\langle\sigma\rangle,v,w) 
 \nonumber \\ &+& {\cal U}(\langle\sigma\rangle,\langle\boldsymbol{\pi}\rangle =0) - {1\over 2} m_v^2\left(v^2 + w^2\right)~,
\end{eqnarray}
together with the condition that $\Omega_{MF}(\langle\sigma\rangle,v,w)$ be minimized with respect to the fields. In the following we use the simplified notation $\langle\sigma\rangle\equiv\sigma$ unless stated otherwise.

The fermionic part with $E = \sqrt{p^2 +M^2(\sigma)}$ and the dynamical nucleon mass $M(\sigma) = g\sigma$ is:
\begin{eqnarray}
\Omega_F = -2\sum_{i=p,n}\int{d^3p\over(2\pi)^3}\left[E +{p^2\over 3E}\sum_{r=\pm 1}n_F(E-r\bar{\mu}_i)\right]\,,\quad 
\label{eq:TpotF}
\end{eqnarray}
where
\begin{equation}
n_F(E\mp\bar{\mu}_i) = \left[\exp\left({E\mp\bar{\mu}_i\over T}\right) +1\right]^{-1}\,,
\end{equation}
with the proton and neutron {\it effective} chemical potentials:
\begin{eqnarray} 
\bar{\mu}_p = \mu_p -g_v v -g_w w~, ~~~~ \bar{\mu}_n = \mu_n -g_v v +g_w w~.
\label{eq:effchempot}
\end{eqnarray}
Mean-field thermodynamics is then determined by the equations for the pressure $P$, the entropy density $s$, the baryon densities $n_i$ and the energy density ${\cal E}$:
\begin{eqnarray}
P&=&-\Omega_{MF}~,~~s = -{\partial\Omega_{MF}\over\partial T}~,~~n_i = -{\partial\Omega_{MF}\over\partial\mu_i}~,\nonumber \\ 
{\cal E} &=& -P +\sum_{i=p,n} \mu_i n_i +Ts ~,
\label{eq:thermodynamics}
\end{eqnarray}
where the mean-field potential is evaluated at its minimum. 
However, this still leaves open the question how to deal with the divergent term proportional to $\int d^3p\,E(p)$ in $\Omega_F$ of Eq.\,(\ref{eq:TpotF}). Many standard MF calculations simply ignored this term. A proper answer to this question has been given in \cite{Skokov2010} which will be pursued in the following.

\subsubsection{Vacuum fluctuations}

The vacuum term
\begin{eqnarray}
\delta\Omega_{vac} = -4\int{d^3p\over(2\pi)^3} E  = -{2\over\pi^2}\int dp\, p^2 \sqrt{p^2+M^2(\sigma)}, \nonumber \\ 
\label{eq:vacterm}
\end{eqnarray}
represents, to lowest order, the one-loop fermionic effective potential, hence the association with fermionic vacuum fluctuations. Dimensional regularisation of Eq.\,(\ref{eq:vacterm}) is done as in Ref.\,\cite{Skokov2010}:
\begin{eqnarray}
\delta\Omega_{vac} = {M^4\over 8\pi^2}\left({2\over 4-d} +{3\over 2} -\gamma_E - \ln{M^2\over4\pi \Lambda^2}\right)~,
\label{eq:dimreg}
\end{eqnarray}
with the Euler-Mascheroni constant $\gamma_E$ and an arbitrary renormalisation scale $\Lambda$. Ultraviolet divergences and irrelevant constants are removed by adding a counter term,
\begin{equation}
\delta{\cal L} = - {M^4\over 8\pi^2}\left({2\over 4-d} +{3\over 2} -\gamma_E + \ln 4\pi\right)~,
\end{equation}
to the Lagrangian. The remaining non-trivial mass dependent logarithmic term can be incorporated in a renormalised bosonic mean-field potential:
\begin{eqnarray}
{\cal U}_B = & &\sum_{n=1}^N {a_n\over 2^n n!} \left(\sigma^2 - f_\pi^2\right)^n - m_\pi^2 f_\pi(\sigma - f_\pi)\nonumber \\&-&{1\over 2}m_v^2\left(v^2 + w^2\right) - {(g\sigma)^4\over 4\pi^2}\ln {g\sigma\over \Lambda}~,
\label{eq:pot1}
\end{eqnarray}
so that the grand canonical potential in this {\it extended} mean-field (EMF) approximation reads:
\begin{eqnarray}
& &~~~~\Omega_{EMF}(T,\mu_p,\mu_n;\sigma,v,w) ={\cal U}_B(\sigma,v,w)  \nonumber \\
&-&{1\over 3\pi^2}\sum_{i=p,n}\int_0^\infty dp\,{p^4\over E}\left[n_F(E-\bar{\mu}_i) +n_F(E+\bar{\mu}_i)\right]\,.\nonumber\\
\label{eq:EMFpot}
\end{eqnarray}
In the vacuum (at $\sigma = f_\pi$) the pressure vanishes. Together with Eq.(\ref{eq:thermodynamics}), it follows that the bosonic potential must be zero at its local minimum: 
\begin{equation}
{\cal U}_B(f_{\pi},0,0) = 0~, ~~{\partial{\cal U}_B\over\partial\sigma}(f_\pi,0,0) = 0~,
\label{eq:vacuumpressure}
\end{equation}
thus $\Lambda = g\,f_\pi$ and the renormalisation scale drops out as it should.
The sigma mass in vacuum is given by the curvature of ${\cal U}_B$ at $\sigma = f_\pi$:
\begin{eqnarray}
m_\sigma^2 = {\partial^2 {\cal U}_B\over\partial\sigma^2} \Big{|}_{\sigma=f_\pi}~.
\label{eq:sigmamass}
\end{eqnarray}
Having constrained the two leading powers in the polynomial expansion of the potential by Eqs.\,(\ref{eq:vacuumpressure}) and (\ref{eq:sigmamass}), one finds:
\begin{eqnarray}
& &{\cal U}_B(\sigma,v,w) = -{g^4\over 4\pi^2}f_\pi^4\ln{\sigma\over f_\pi} - m_\pi^2 f_\pi(\sigma - f_\pi)\nonumber \\
&+&{1\over 2}\left[m_\pi^2 + {g^4\over 4\pi^2}f_\pi^2\left(1-4\ln{\sigma\over f_\pi}\right)\right]\left(\sigma^2 - f_\pi^2\right)\nonumber \\
&+&{1\over 8}\left[{m_\sigma^2-m_\pi^2\over f_\pi^2} + {g^4\over 2\pi^2}\left(3-4\ln{\sigma\over f_\pi}\right)\right]\left(\sigma^2 - f_\pi^2\right)^2\nonumber \\
&+&\sum_{n=3}^N {a_n\over 2^n n!}\left(\sigma^2 - f_\pi^2\right)^n - {1\over 2}m_v^2(v^2+w^2)~.
\label{eq:pot2}
\end{eqnarray}

In the following we shall distinguish between 
\begin{equation}
\Omega^{(0)}_{MF} \equiv \Omega_{EMF} + {(g\sigma)^4\over 4\pi^2}\ln{\sigma\over f_\pi}~,
\label{eq:MFpot}
\end{equation}
(MF without logarithmic vacuum terms), and $\Omega_{EMF}$ (with inclusion of vacuum terms).
The minimisation of either $\Omega^{(0)}_{MF}$ or $\Omega_{EMF}$ at fixed chemical potentials $\mu_p$ and $\mu_n$ yields the vector field equations: 
\begin{eqnarray}
 v &=& {g_v\over m_v^2}\left[n_p + n_n\right]~,\nonumber \\
w &=&  {g_w\over m_v^2}\left[n_p - n_n\right]~, 
\label{eq:vectorfields}
\end{eqnarray}
and the equation for the scalar field:
\begin{equation}
{\partial{\cal U}_B\over\partial\sigma} = -g\,n_s~.
\end{equation}
These field equations are solved self-consistently at given temperature and chemical potentials, with the proton and neutron densities:
\begin{eqnarray}
& &n_{p,n}(T,\mu_p,\mu_n;\sigma,v,w) = -{\partial\Omega\over\partial\mu_{p,n}}  \nonumber \\
&=& {1\over\pi^2}\int_0^\infty dp\,p^2\left[n_F(E-\bar{\mu}_{p,n}) - n_F(E+\bar{\mu}_{p,n})\right]~,\quad
\end{eqnarray}
and with the scalar density:
\begin{eqnarray}
& &~~n_s(T,\mu_p,\mu_n;\sigma,v,w) = {\partial\Omega\over\partial M}  \nonumber \\
&=&\sum_{i=p,n}{1\over\pi^2}\int_0^\infty dp{p^2M\over E}\left[n_F(E-\bar{\mu}_i)+ n_F(E+\bar{\mu}_i)\right]~.\nonumber\\
\end{eqnarray}
Note that the vector field contributions to ${\cal U}_B$ involve the masses and coupling constants only in the combinations
\begin{equation}
G_v =  {g_v^2\over m_v^2}~~,~~~G_w = {g_w^2\over m_v^2}~.
\end{equation}
With a chosen maximum order $N=4$ in the polynomial expansion of the potential, five remaining parameters, \\ $(a_3,a_4,m_\sigma,G_v,G_w)$, need to be fixed by reproducing empirical nuclear bulk data, such as ground state properties of nuclear matter and characteristics of the liquid-gas phase transition.

\subsubsection{Parameter fixing: nuclear thermodynamics}

Results of calculations without the logarithmic vacuum term (MF) and including this term (EMF) will be compared in a subsequent section, especially with respect to the role of vacuum fluctuations in determining the chiral order parameter $\langle\sigma\rangle$ as a function of chemical potentials. In order to prepare the ground for such an investigation, a necessary prerequisite is an optimal reproduction of well established nuclear phenomenology.

\begin{figure*}[t]
\begin{center}
\includegraphics[height=58mm,angle=-00]{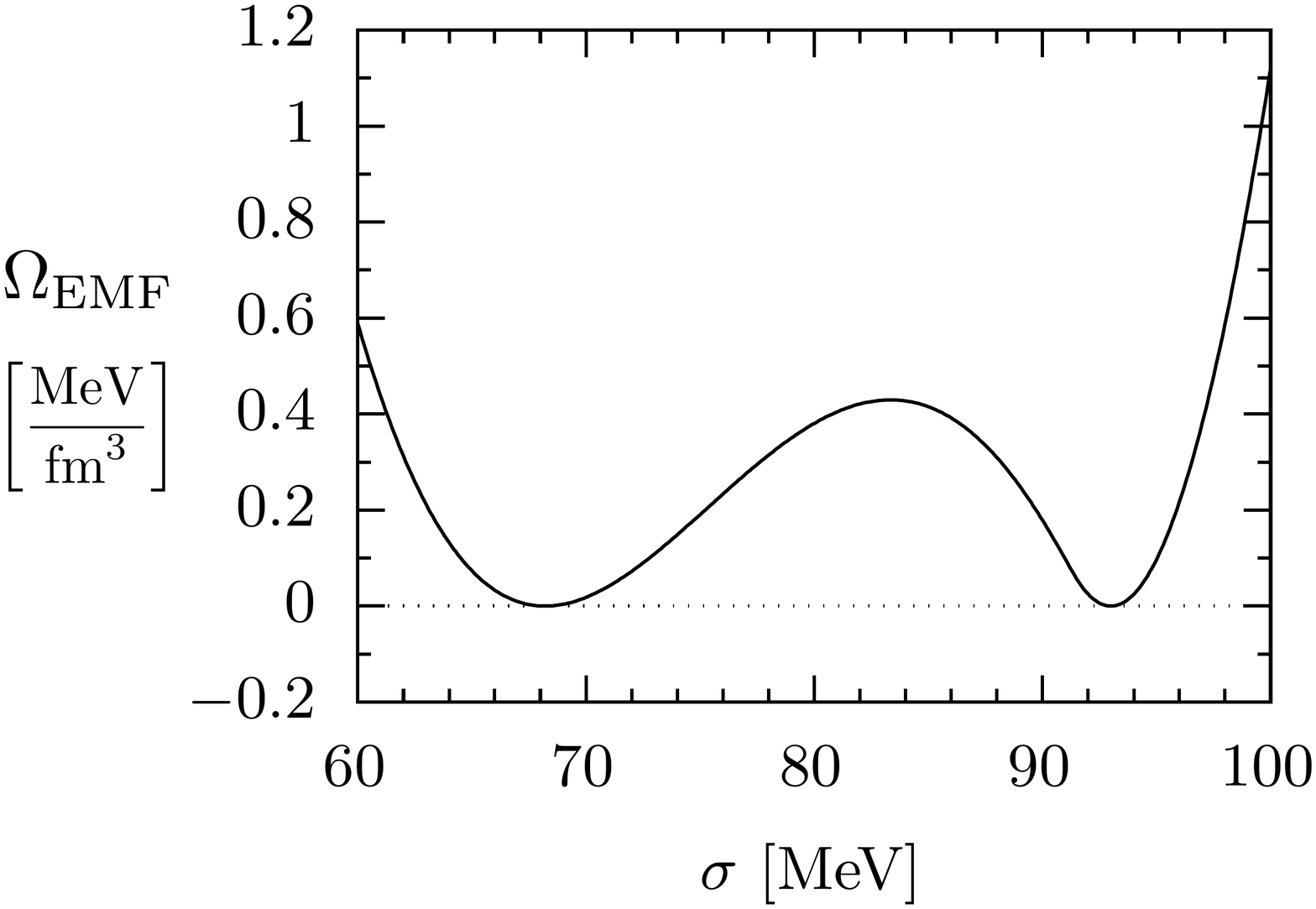}
\caption{Thermodynamic potential of the ChNM model for symmetric nuclear matter at temperature $T = 0$ and baryon chemical potential $\mu = 923$ MeV, in the extended mean-field (EMF) approximation, as function of the expectation value of the scalar (sigma) field. The two minima correspond to the vacuum ($\langle\sigma\rangle_{vac} = f_\pi \simeq 93$ MeV) and to the nuclear matter ground state ($\langle\sigma\rangle_0 = 0.74\,f_\pi \simeq 68$ MeV). }
\label{fig:1}
\end{center}
\end{figure*}

We use the following criteria:

i) Ground state energy, $E_0/A = -16$ MeV, at a saturation density $n_0 = 0.16$ fm$^{-3}$, and symmetry energy, $S_0 = 32$ MeV, of nuclear matter \cite{footnote2};

ii) Nuclear first-order liquid-gas phase transition: at zero temperature the potential has two degenerate minima in $\sigma$ with vanishing pressure corresponding to the vacuum and to the ground state of symmetric nuclear matter;

iii) Empirical parameters of the liquid-gas phase transition at its critical point \cite{Elliot2013}: critical temperature $T_{\textrm{crit}} = 17.9\pm 0.4$ MeV, pressure $P_{\textrm{crit}} = 0.31\pm 0.07$ MeV/fm$^3$ and baryon density $n_{\textrm{crit}} = 0.06\pm 0.01$ fm$^{-3}$.

In addition, the following conditions have been imposed:

iv) The empirical value of the nuclear surface tension is $\Sigma =  1.08 \pm 0.06\,\textrm{MeV/fm}^2$ \cite{Hua1999};

v) The Landau effective mass of the nucleon quasiparticles at the Fermi surface of nuclear matter should be in the range $M^*_L = 0.7 - 0.8\,M_N$ \cite{Glendenning1997}.\\  

It turns out that the following set of EMF parameters optimally fulfils these criteria:
\begin{eqnarray}
m_\sigma &=& 617.6\, \textrm{MeV}~,~G_v = 5.88\,\textrm{fm}^2~,~G_w=0.97\,\textrm{fm}^2~,\nonumber \\
a_3 &=& 2.16\cdot 10^{-1}\,\textrm{MeV}^{-2}~,~a_4 = -5.29\cdot 10^{-5}\,\textrm{MeV}^{-4}~,\nonumber\\
\label{eq:EMFparameters}
\end{eqnarray}
together with the scalar-pseudoscalar Yukawa coupling $g = 10.1$. The chemical potential in symmetric nuclear matter at $T=0$ equilibrium is:
\begin{equation}
\mu_0 = M_N + {E_0\over A} = 923\,\textrm{MeV}~.
\end{equation}
Fig.\,\ref{fig:1} shows the EMF potential $\Omega_{EMF}$ at zero temperature and chemical potential $\mu_0 = 923$ MeV as a function of the scalar mean field $\sigma$. The two degenerate minima correspond to the vacuum with $\sigma = \langle\sigma\rangle_{vac} = f_\pi$ and to the ground state of symmetric nuclear matter with $\sigma = \langle\sigma\rangle_0 = 68.2$ MeV $= 0.74\,f_\pi$.

This set of parameters gives the critical liquid-gas transition parameters for symmetric nuclear matter:
\begin{eqnarray}
T_{\textrm{crit}}&=& 17.5\,\textrm{MeV}~,~P_{\textrm{crit}} = 0.33\,\textrm{MeV/fm}^3~,\nonumber\\
n_{\textrm{crit}}&=& 0.06\,\textrm{fm}^{-3}~,~\mu_{\textrm{crit}} = 908\,\textrm{MeV}~,
\end{eqnarray}
in agreement with the empirical values. 

Following \cite{Coleman1977} one can estimate the nuclear surface tension as a measure of the potential barrier thickness between the two minima:
\begin{equation}
\Sigma = \int_{\langle\sigma\rangle_0}^{f_\pi} d\sigma\sqrt{2\Omega_{EMF}(\sigma)} = 1.1\,\textrm{MeV/fm}^2~,
\end{equation}
which constrains the $\sigma$ mass $m_\sigma$ as given in (\ref{eq:EMFparameters}). This is in agreement with the empirical value and the surface tension deduced in \cite{Berges2003} from the mass formula of the nuclear droplet model.

\begin{figure*}[t]
\begin{center}
\includegraphics[height=90mm,angle=-00]{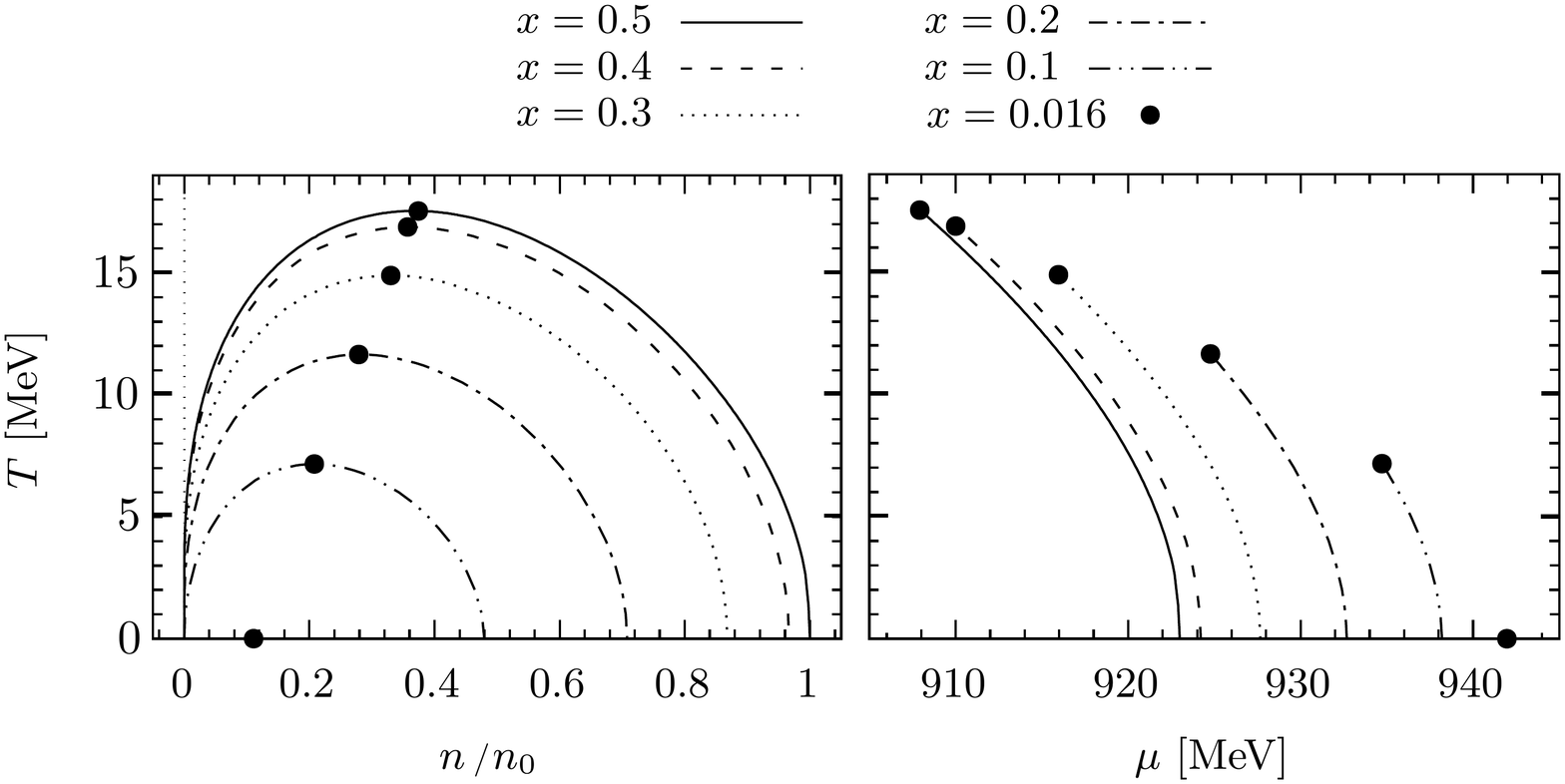}
\caption{Pattern of the liquid-gas first-order phase transition in nuclear matter. The systematic evolution of the liquid-gas coexistence region in a $(T,n)$ diagram, and of the phase transition line in a $(T,\mu)$ diagram, are displayed for various values of the proton fraction, $x = Z/A$, from symmetric nuclear matter to neutron-rich matter.}
\label{fig:2}
\end{center}
\end{figure*}

The Landau effective mass of the nucleon quasiparticles at the Fermi surface of nuclear matter (with Fermi momentum $p_F = 263$ MeV) becomes: 
\begin{equation}
M^*_L = \sqrt{p_F^2 +(g\,\langle\sigma\rangle_0)^2} = \mu_0 - G_v\,n_0 = 0.79\,M_N~.
\end{equation}
The compression modulus, $K = 9n(dn/d\mu)^{-1}$, at $n=n_0$ comes out as $K = 282$ MeV, slightly larger than the empirical range $K = 240\pm 20$ MeV \cite{Gil2020} but still acceptable.

As further checks we examine the systematic evolution of the liquid-gas phase transition pattern for asymmetric nuclear matter. In order to properly analyse, at constant proton fraction $x = Z/A$, the coexistence region of liquid and gas phases (with different $\mu_p$ and $\mu_n$ in each) and the corresponding transition curves in the $T-\mu$ diagram, an average nucleonic chemical potential
\begin{equation}
\mu = x\mu_p + (1-x)\mu_n ~,
\end{equation}
needs to be introduced. Results are displayed in Fig.\,\ref{fig:2}. 
As can be seen, going from symmetric nuclear matter to increasingly neutron rich matter, the critical point 
moves to lower temperatures and lower densities until the liquid-gas transition disappears altogether at $x\simeq 0.02$. 

With the isovector-vector coupling $G_w \simeq 1$ fm$^2$ as given in (\ref{eq:EMFparameters}), pure neutron matter has an energy per particle $E/A = +16$ MeV at $n = n_0$ so that the symmetry energy, $S(n_0) = (E(n_0,x=0)-E(n_0,x=0.5))/A = 32$ MeV, is within its empirical range. The symmetry energy at about twice the density of nuclear matter was extracted from Au+Au collisions at an energy of 400 MeV per nucleon \cite{Russotto2016}: $S(n = 2n_0) = (55\pm5)$ MeV. We find a slightly larger value, $S(2n_0) = 62$ MeV.

\subsection{Functional renormalisation group}

The $\Omega_{EMF}$ version of the grand canonical potential provides a good description of baryonic matter at densities around $n \sim n_0 = 0.16\,\textrm{fm}^{-3}$. Omitting vacuum fluctuations, the simple MF version $\Omega^{(0)}_{MF}$ works similarly well at low densities, with marginal readjustment of input parameters \cite{Floerchinger2012}. However, differences between $\Omega_{EMF}$ and $\Omega^{(0)}_{MF}$, i.e. the vacuum terms, become qualitatively important as one proceeds to higher baryon densities, as we shall see. Still these vacuum fluctuations do not cover many other ``soft" degrees of freedom, such as important loop effects involving chiral bosons and nucleons. A method to deal with this broader range of fluctuations beyond MF is the functional renormalisation group (FRG).

The FRG scheme, applied here to the ChNM model, proceeds as follows (we refer to \cite{Drews2014,Drews2015,Drews2017} for more detailed derivations and discussions). An effective action depending on a renormalisation scale $k$ is introduced:
\begin{eqnarray}
\Gamma_k &=&\int^{1/T}_0 dx_4\int{d^3x} \Bigl\{\bar{\Psi}\left[\gamma_\mu\partial_\mu + g(\sigma + i\gamma_5\,\boldsymbol{\tau\cdot\pi})\right]\Psi \nonumber \\
&+&\Psi^\dagger\left(\boldsymbol{\mu} - g_v\,v - g_w\, \tau_3\,w  \right)\Psi + {\frac12}\left(\partial_\mu \sigma \partial_\mu \sigma + \partial_\mu \boldsymbol{\pi}\cdot\partial_\mu \boldsymbol{\pi}\right) \nonumber\\
&+&{\cal U}_k(T,\mu_p,\mu_n;\sigma, \boldsymbol{\pi}, v,w)\Bigr\} ~,
\label{eq:k-action}
\end{eqnarray}
with
\begin{eqnarray}
\boldsymbol{\mu} = \begin{pmatrix}
\mu_p & \\
& \mu_n
\end{pmatrix}~.
\end{eqnarray} 
The action $\Gamma_k$ is initialized at an ultraviolet (UV) scale of order 1 GeV, $k_{UV} \sim \Lambda_\chi = 4\pi f_\pi$,  the characteristic scale of spontaneous chiral symmetry breaking \cite{footnote3}. Starting from $\Gamma_{k=k_{UV}}$ the flow of $\Gamma_k$ is determined in such a way that it interpolates between the primary UV action and a suitable quantum effective action $\Gamma_{\rm eff}=\Gamma_{k=0}$ in the infrared (IR) limit, $k\rightarrow 0$. The evolution of $\Gamma_k$ as a function of $k$ is given by Wetterich's flow equation \cite{Wetterich1993}, schematically written as 
\begin{align}\label{eq:Wetterich}
	\begin{aligned}
		k\,\frac{\partial\Gamma_k}{\partial k}=&
		\begin{aligned}
			\includegraphics[width=0.1\textwidth]{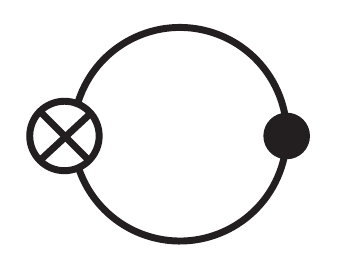}
		\end{aligned} \\ =&\frac 12 {\rm Tr}\left[k\frac{\partial R_k}{\partial k}\cdot\Big(\Gamma_k^{(2)}[\Phi]+R_k\Big)^{-1}\right]~~.
	\end{aligned}
\end{align}
The trace Tr stands for all relevant sums and integrations including a sum over the bosonic and fermionic subspaces, where the fermionic contribution comes with an additional minus sign and $\Phi$ denotes the set of all active fermion and boson fields.  A scale regulator, $R_k(p)$, is introduced such that $\Gamma_k$ contains all fluctuations with momenta $p^2 \gtrsim k^2$, whereas fluctuations with $p^2 \lesssim k^2$ are suppressed. 
%In this way $\Gamma_k$ interpolates between the bare action in the UV and the full effective action in the IR limit.
In practice the optimized $k$-regulator for bosons, $R_k(\boldsymbol{p}) = (k^2-\boldsymbol{p}^2)\theta(k^2-\boldsymbol{p}^2)$ is used, as in Refs.\cite{Litim2001,Litim2006,Blaizot2007}, together with a corresponding regulator for Dirac particles. The matrix $\Gamma_k^{(2)}$ involves 2nd functional derivatives of the effective action with respect to chiral and nucleon fields. It collects the {\it full} inverse propagators of all particles involved. In the pictorial illustration of the flow equation (\ref{eq:Wetterich}) these full propagators are marked by the dot on the loop line while the $k$-regulator is symbolized by the crossed circle.

At this point it is assumed that the Yukawa couplings $g, g_v$ and $g_w$ are not scale dependent by themselves. Furthermore the vector fields, $v$ and $w$, are treated again as mean fields, their masses being sufficiently large so that fluctuations of these fields can be neglected. So the explicit $k$-dependence rests primarily in the effective potential:
\begin{equation}
{\cal U}_k = {\cal U}_k^{(0)}(\chi) - m_\pi^2 f_\pi(\sigma - f_\pi) - {1\over 2}m_v^2(v^2+w^2)~.
\end{equation}
It includes the scale dependent chirally symmetric piece,
\begin{equation}
{\cal U}_k^{(0)}(\chi) = \sum_{n=0}^N {a_n(k)\over n!}\left(\chi - \chi_0\right)^n~,
\label{eq:chipot}
\end{equation}
with $\chi = {1\over 2}\left(\sigma^2 + \boldsymbol{\pi}^2\right)$ and the vacuum expectation value $\chi_0 =  {1\over 2}\langle\sigma\rangle_{vac}^2 = {1\over 2} f_\pi^2$. The coeficients $a_n(k; T,\bar{\mu}_p,\bar{\mu}_n)$ are functions of temperature and effective chemical potentials, not displayed in Eq.\,(\ref{eq:chipot}) for simplicity. Note that the logarithmic vacuum term in Eqs.\,(\ref{eq:pot1}) and (\ref{eq:pot2}) of the EMF potential must not be added here because its non-perturbative extension, encoded in ${\cal U}_k^{(0)}(\chi)$, is generated together with other fluctuations in the non-perturbative FRG scheme. 

For the treatment of a dense and thermal medium with inclusion of fluctuations it is useful to compute the flow of the difference between the effective action at given values of temperature and chemical potential, $\Gamma_k(T,\mu)$, as compared to the potential at a reference point for which we choose either the vacuum, $\Gamma_k(0,0)$, or equilibrium nuclear matter at zero temperature, $\Gamma_k(0,\mu_0)$ with $\mu_0 = M_N + E_0/A = 923$ MeV. The latter choice is favoured in the case of the ChNM model. The flow of the difference, $\bar\Gamma_k=\Gamma_k(T,\mu)-\Gamma_k(0,\mu_0)$, satisfies the FRG equation
\begin{align}
	\begin{aligned}
		\frac{k\,\partial\bar \Gamma_k}{\partial k}(T,\mu)&=
		\begin{aligned}
			\hspace{-.1cm}
			\vspace{1cm}
			\includegraphics[width=0.1\textwidth]{wetterich_fermion}
		\end{aligned} \vspace{-1cm}\Bigg|_{T,\mu}-
		\begin{aligned}
			\hspace{-.1cm}
			\vspace{1cm}
			\includegraphics[width=0.1\textwidth]{wetterich_fermion}
		\end{aligned} \Bigg|_{\begin{subarray}{l} T=0 \\ \mu=\mu_0 \end{subarray}}.
	\end{aligned}
\end{align}
Note that the $k$-dependent effective action in Eq. (\ref{eq:k-action}) is treated in leading order of the derivative expansion and we work in the local potential approximation, neglecting (small) wave function renormalisation effects on the chiral boson fields and possible higher order derivative couplings. With these assumptions and truncations the flow equation is, of course, not exact any more. The dependence of the nucleon mass on temperature and chemical potential scales with that of the in-medium pion decay constant, $f_\pi^*(T,\mu) = \langle\sigma\rangle(T,\mu)$, which acts as a chiral order parameter.

For homogeneous fields the Euclidean volume in the action (\ref{eq:k-action}) factors out. The only remaining scale-dependent part is the chirally invariant potential, so that
\begin{equation}
{\partial\Gamma_k\over\partial k} = {V\over T}\,{\partial{\cal U}^{(0)}_k\over\partial k}(T,\mu_p,\mu_n;\chi,v,w)\,.
\end{equation}
Following \cite{Drews2015,Drews2017} the flow equation for ${\cal U}^{(0)}_k$ becomes
\begin{eqnarray}
{\partial{\cal U}_k^{(0)}\over\partial k} &=& {k^4\over 12\pi^2}\Biggl\{{1\over E_\sigma}\left[1+2n_B(E_\sigma)\right]+{3\over E_\pi}\left[1+2n_B(E_\pi)\right]\nonumber\\
 &-&{4\over E_N}\sum_{i=p,n}\Big[1- n_F(E_N-\bar{\mu}_i)-n_F(E_N+\bar{\mu}_i)\Big]\Biggr\}\,.\nonumber\\
\label{eq:potflow}
\end{eqnarray}
Here,
\begin{eqnarray}
E_N^2 &=& k^2 + 2g^2\chi~,~~~E_\pi^2 = k^2 + {\partial{\cal U}^{(0)}_k\over\partial\chi}\,,\nonumber \\
E_\sigma^2 &=& k^2 + {\partial{\cal U}^{(0)}_k\over\partial\chi} + 2\chi{\partial^2{\cal U}^{(0)}_k\over\partial\chi^2}\,,
\label{eq:energies}
\end{eqnarray}
and $n_{F,B}(E) = \left[\exp(E/T) \pm 1\right]^{-1}$. 
Eq.\,(\ref{eq:potflow}) is then a set of coupled differential equations for the coefficients $a_n(k)$ in Eq,\,(\ref{eq:chipot}), which are solved using a grid method. The leading coefficients, $a_1$ and $a_2$, can be expressed at $k = k_{UV}$ by the pion and sigma masses 
\begin{eqnarray}
a_1(k_{UV}) = m_{\pi}^2~,~~~ a_2(k_{UV})=\frac{m_{\sigma}^2(k_{UV}) - m_\pi^2}{2f_\pi^2}~.
\end{eqnarray}
Because the pion mass is fixed to its physical value for all $k$ by the explicit chiral symmetry breaking term, this leaves $m_\sigma(k_{UV})$, $a_3(k_{UV})$ and $a_4(k_{UV})$ as remaining parameters to be determined.

The vector fields $v$ and $w$ are once again treated as background mean fields, but their implicit $k$-scale dependence along the flow path from the UV initialization to the IR limit, $k\rightarrow 0$, needs now to be taken into account to reduce computational cost. For this purpose the effective potential ${\cal U}_k$ is minimized with respect to the vector fields at every value of $k$. The resulting fields $\bar{v}_k$ and  $\bar{w}_k$ satisfy the equations:
\begin{eqnarray}
& &~~~~~~~~~~~~~~~g_v\,\bar{v}_k = {G_v\over 3\pi^2}\int_k^{k_{UV}}dp\, {p^4\over E_N} \nonumber\\
&\times& \sum_{r=\pm1}{\partial\over\partial\mu}\Bigl\{n_F(E_N - r\mu)\large|_{\mu=\bar{\mu}_p}+n_F(E_N - r\mu)\large|_{\mu=\bar{\mu}_n}\Bigr\}~,\nonumber\\
& &~~~~~~~~~~~~~~~g_w\,\bar{w}_k = {G_w\over 3\pi^2}\int_k^{k_{UV}}dp \,{p^4\over E_N} \nonumber\\
&\times& \sum_{r=\pm1}{\partial\over\partial\mu}\Bigl\{n_F(E_N - r\mu)\large|_{\mu=\bar{\mu}_p}-n_F(E_N - r\mu)\large|_{\mu=\bar{\mu}_n}\Bigr\}~,\nonumber\\
\end{eqnarray}
which replace the mean-field equations (\ref{eq:vectorfields}). These vector fields depend in addition on the chiral field $\chi$ through the nucleon energy $E_N = \sqrt{p^2 + 2g^2\chi}$ in Eq.\,(\ref{eq:energies}). The proton and neutron effective chemical potentials, $\bar{\mu}_p$ and $\bar{\mu}_n$, have the same form as in Eq.\,(\ref{eq:effchempot}) but they are now implicitly $k$-dependent. Finally the IR limit $k\rightarrow 0$ is taken \cite{footnote4} and the potential is minimized with respect to the $\chi$ field. This defines the grand canonical potential:
\begin{equation}
\Omega_{FRG}(T,\mu_p,\mu_n) = {\cal U}_{k=0}(T,\mu_p,\mu_n;\bar{\sigma},\bar{v}_{k=0},\bar{w}_{k=0})~,
\end{equation}
where the chiral field at the minimum is denoted $\bar{\chi} = {1\over 2}\bar{\sigma}^2$. The assumed absence of a pion condensate means that the expectation value of the pion field vanishes in the IR limit. Nonetheless the FRG approach treats fluctuations of the pion field explicitly, in contrast to the mean-field approximation.

\begin{figure*}[t]
\begin{center}
\includegraphics[height=60mm,angle=-00]{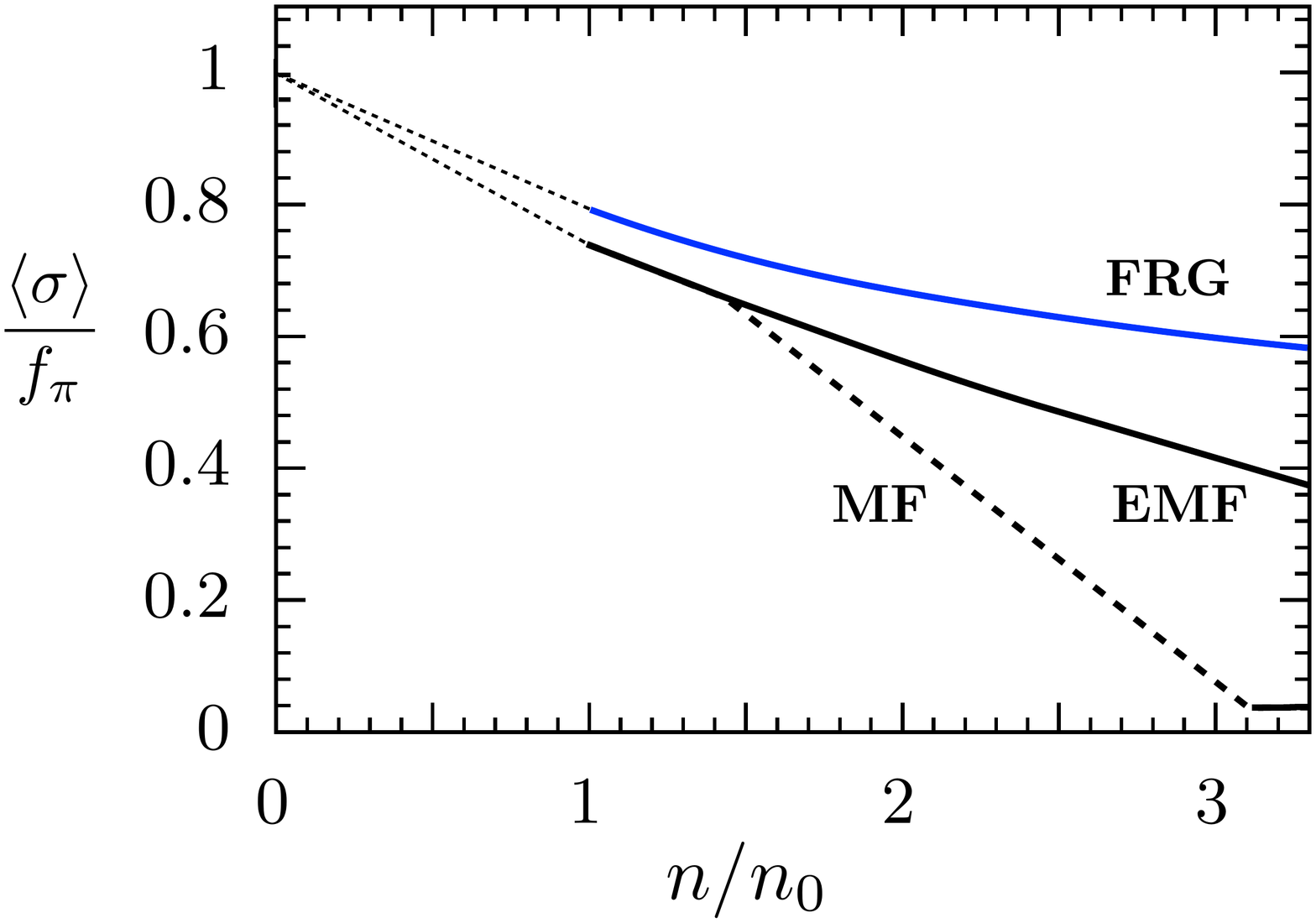}
\caption{Chiral order parameters in symmetric nuclear matter at temperature $T=0$ as a function of baryon density. $n$ in units of nuclear ground state equilibrium density, $n_0 = 0.16$ fm$^{-3}$. Dotted lines: liquid-gas phase transition; dashed line: first-order chiral phase transition. Plotted are the results from basic mean-field (MF) and extended mean-field approximations (EMF, with inclusion of vacuum fluctuations). Also shown is the curve resulting from a functional renormalisation group (FRG) computation based on the same ChNM model \cite{Drews2015}.  }
\label{fig:3}
\end{center}
\end{figure*}

The set of FRG input parameters at the scale $k_{UV} = 1.4$ GeV, optimizing the comparison with nuclear data, have been determined in Ref.\,\cite{Drews2015}:
\begin{eqnarray}
m_\sigma &=& 770\, \textrm{MeV}~,~G_v = 4.04\,\textrm{fm}^2~,~G_w=1.12\,\textrm{fm}^2~,\nonumber \\
a_3 &=& 5.55\cdot 10^{-3}\,\textrm{MeV}^{-2}~,~~a_4 = 8.38\cdot 10^{-5}\,\textrm{MeV}^{-4}~.\nonumber\\
\label{eq:FRGparameters}
\end{eqnarray}
The changes of these parameters in comparison to those of the EMF scheme, Eq.\,(\ref{eq:EMFparameters}), are of some significance. Consider for example the input $\sigma$ boson mass which has a large UV starting value, $m_\sigma \simeq 0.8$ GeV, in the FRG approach. The dynamical evolution towards the infrared scale, $k\rightarrow 0$, results in a strong downward shift of this mass. Indeed the FRG sigma mass in vacuum calculated at the minimum of the effective potential in the IR limit becomes
\begin{equation}
m_\sigma^{IR} = \sqrt{{\cal U}'_{k=0}(\chi_0) + 2\chi_0\,{\cal U}''_{k=0}(\chi_0)} \simeq 0.6\,\textrm{GeV}\nonumber
\end{equation}
at $\chi_0 = f_\pi^2/2$. This value is close to $m_\sigma$ in the MF or EMF schemes which do not handle fluctuations explicitly and therefore compensate for this by the choice of a lower input sigma mass. One might be reminded here of the relatively low mass, $m_\sigma \simeq 0.44$ GeV, deduced from the pole in the isoscalar s-wave $\pi\pi$ scattering amplitude \cite{Caprini2006}.

The isoscalar vector coupling strength, $G_v$, represents the effects of the repulsive short-range core of the nucleon-nucleon interaction. Its mean-field value is considerably larger than the one required using FRG. In the FRG scheme part of the short-range repulsion is generated by high-momentum fluctuation effects with inclusion of Pauli principle corrections. The MF and EMF approaches do not incorporate such mechanisms explicitly and must therefore compensate for their absence by an increased $G_v$.

\section{Phase structure and chiral order parameters}

We are now prepared to enter the detailed study of phases of baryonic matter, with special focus on the chiral order parameter and its dependence on fluctuations. The chiral order parameter is identified with the pion decay constant, $f_\pi$, and its behaviour as a function of temperature and baryon density or chemical potential. This decay constant is defined by the matrix element of the time component of the QCD axial current (the axial density) connecting the vacuum with a Nambu-Goldstone pion at rest: $\langle 0|\psi^\dagger\gamma_5\tau_a\psi |\pi_b\rangle = im_\pi f_\pi\delta_{ab}$. It is related to the chiral (quark) condensate, to leading order in the $u$ and $d$ current quark masses, by $m_\pi^2 f_\pi^2 = {1\over 2}(m_u+m_d)\langle\bar{q}q\rangle$. In the context of the ChNM model the chiral order parameter is the expectation value of the scalar field, $\langle\sigma\rangle(T,\mu)$, normalized to $\langle\sigma\rangle(T=0,\mu=0) = \langle\sigma\rangle_{vac} = f_\pi$ in the vacuum. At non-zero temperature and chemical potential the vacuum is replaced by the dense thermal medium. In the following our focus will be on $\langle\sigma\rangle$ at $T = 0$ as a function of baryon density or chemical potential.

\begin{figure*}[t]
\begin{center}
\includegraphics[height=65mm,angle=-00]{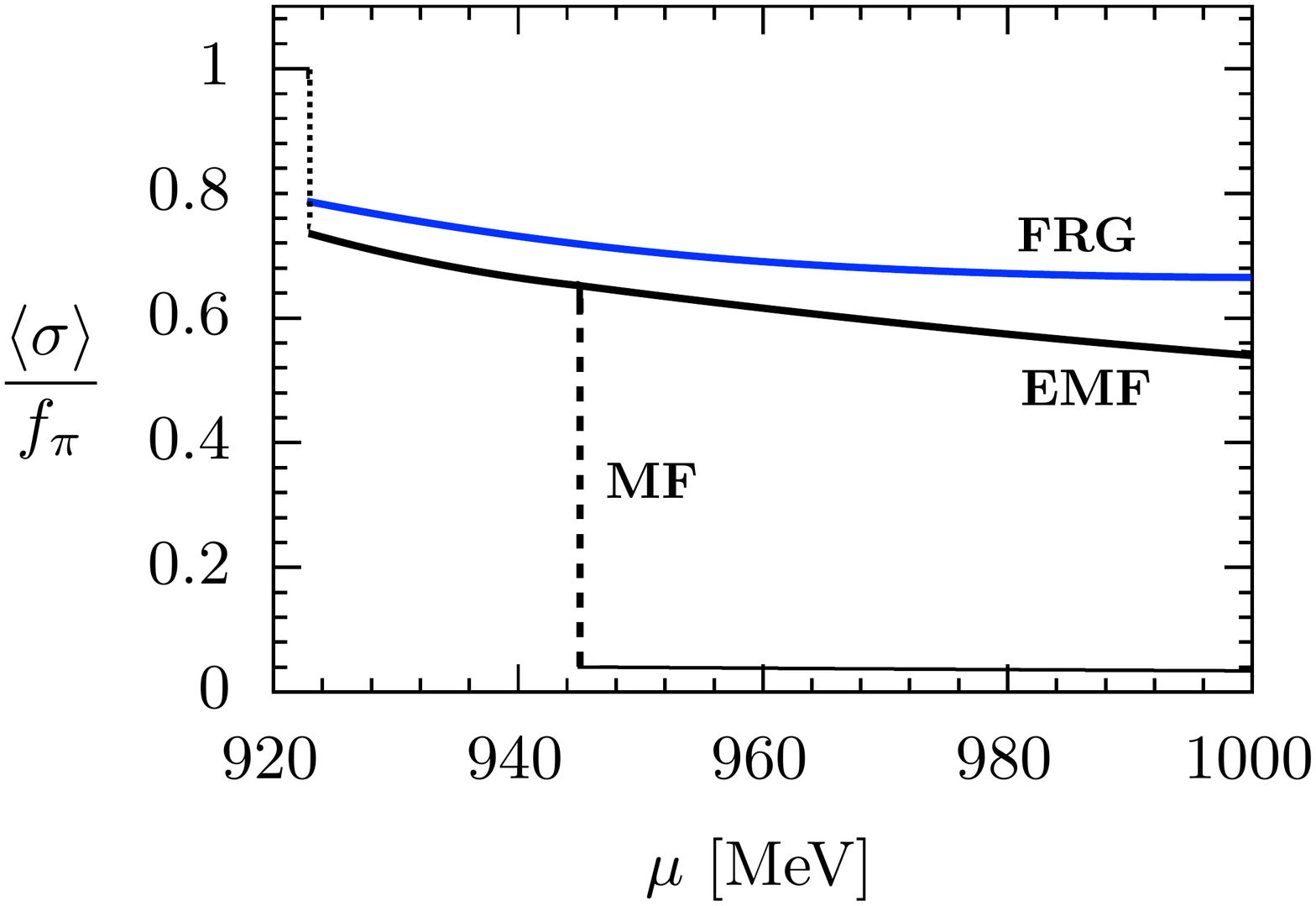}
\caption{Chiral order parameters in symmetric nuclear matter at $T=0$ as a function of baryon chemical potential $\mu$. Legends are the same as in Fig.\,\ref{fig:3}.}
\label{fig:4}
\end{center}
\end{figure*}

\begin{figure*}[t]
\begin{center}
\includegraphics[height=60mm,angle=-00]{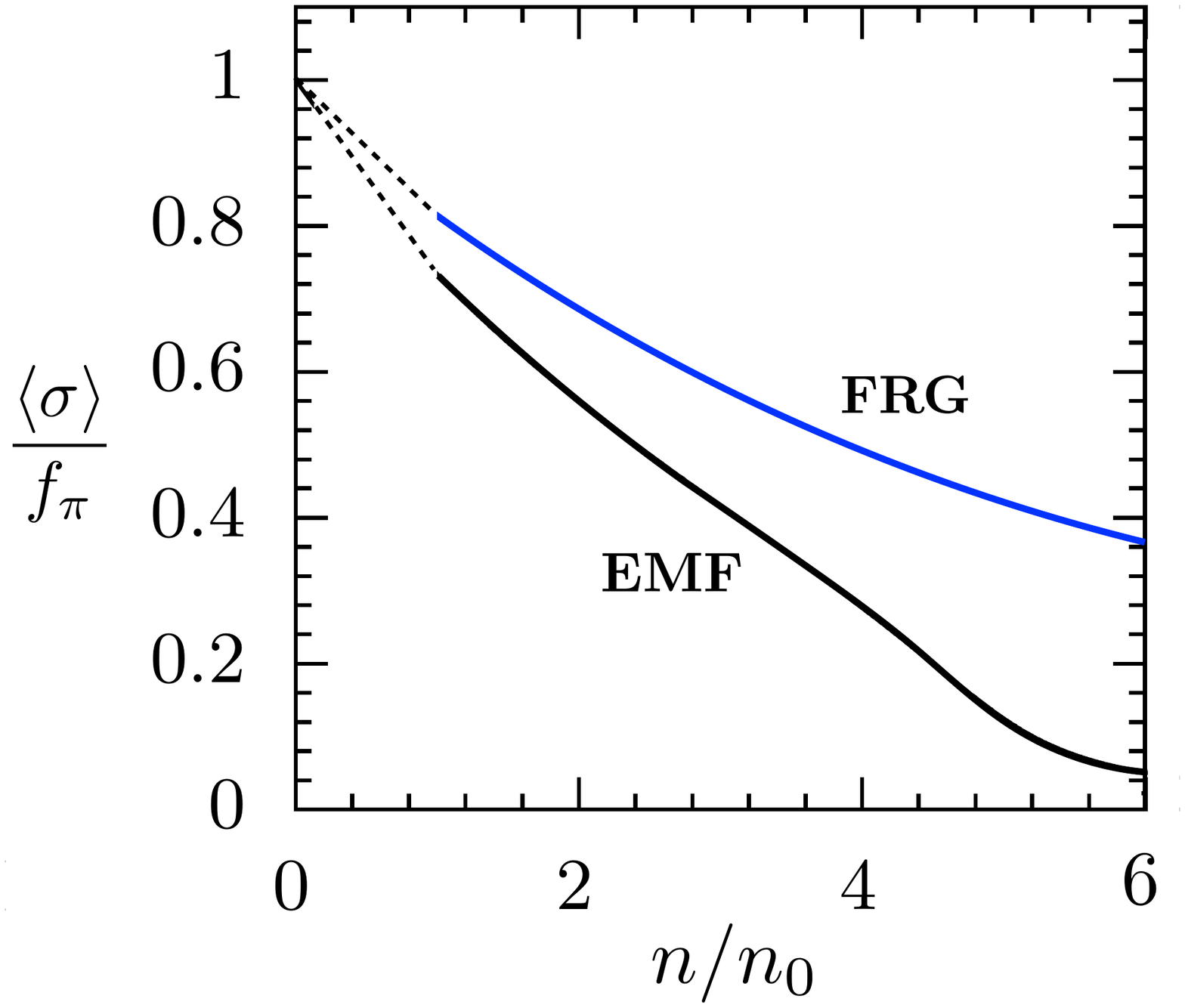}
\caption{Chiral order parameters in symmetric nuclear matter using the ChNM model as described in the text. Shown is in particular the behaviour at high baryon densities (in units of $n_0 = 0.16$ fm$^{-3}$) for the EMF and FRG scenarios. Dotted lines at densities $n\le n_0$ describe the first-order liquid-gas phase transition.}
\label{fig:5}
\end{center}
\end{figure*}

\subsection{Symmetric nuclear matter}

Reproducing the thermodynamics of the liquid-gas first-order phase transition has been one of the basic criteria for a realistic initialization of the ChNM model and further extrapolations. This first-order phase transition leaves its signature as well in the chiral order parameter for symmetric nuclear matter, as shown in Fig.\,\ref{fig:3} as a function of baryon density. A first instructive step is the comparison between the MF case (mean-field approximation without vacuum fluctuation terms), using $\Omega^{(0)}_{MF}$ of Eq.\,(\ref{eq:MFpot}), and its EMF extension (with inclusion of vacuum terms), using $\Omega_{EMF}$ of Eq.\,(\ref{eq:EMFpot}). The MF approximation would suggest a first-order chiral phase transition starting already at a density as low as $n\simeq 1.5\,n_0$, with a coexistence region extending up to $n\simeq 3\,n_0$ at which chiral symmetry is fully restored. This is clearly an unphysical situation. Such a qualitative phase change would already have been noticeable in the empirical nuclear phenomenology and in heavy-ion collisions. With inclusion of vacuum terms as a minimally added condition, this chiral first-order transition at low density disappears indeed and shifts to high densities far beyond $3\,n_0$. 

Fig.\,\ref{fig:4} shows the corresponding picture as a function of baryon chemical potential. A sudden jump from the vacuum expectation value $\langle\sigma\rangle = f_\pi$ takes place at $\mu_0 = 923$ MeV due to the liquid-gas phase transition. In the MF case a chiral first-order phase transition would appear not much further up, at $\mu_c = 945$ MeV, above which the nucleon mass, $M = g\langle\sigma\rangle$, would vanish. In the extended EMF scheme with inclusion of the vacuum term (zero point energy density), this is evidently not the case any more. The chiral order parameter as well as the nucleon mass stays non-zero over a much wider range of chemical potentials.

\begin{figure*}[t]
\begin{center}
\includegraphics[height=60mm,angle=-00]{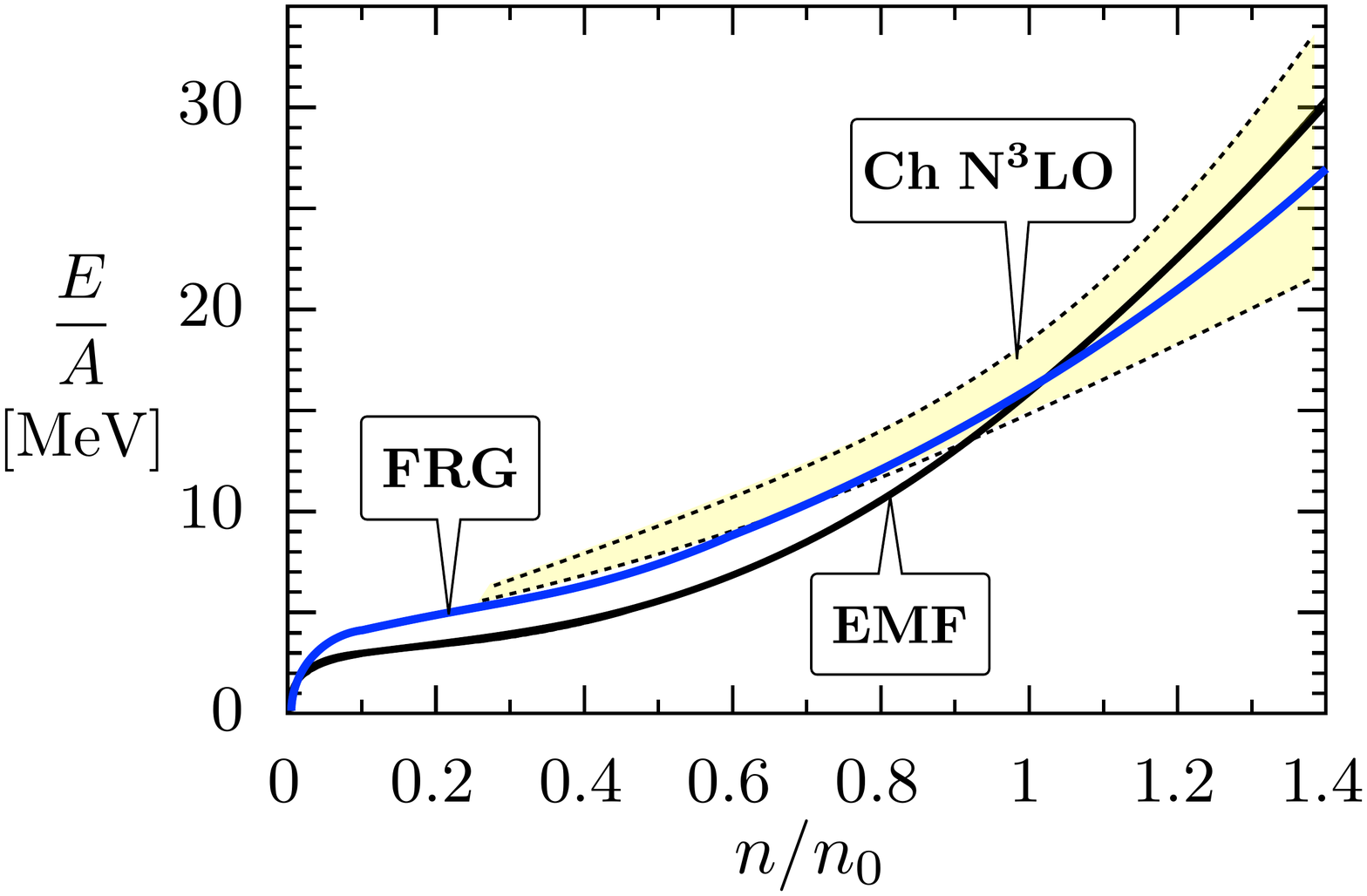}
\caption{Energy per particle of pure neutron matter as a function of neutron density. Curves show results of  calculations based on the ChNM model: EMF as described in the text and FRG with reference to \cite{Drews2015}. The light-shaded band shows for comparison the $E/A$ (including uncertainties) obtained with a chiral $N^3LO$ nucleon-nucleon interaction plus three- and four-body interaction terms \cite{Drischler2019}. }
\label{fig:6}
\end{center}
\end{figure*}

\begin{figure*}[t]
\begin{center}
\includegraphics[height=60mm,angle=-00]{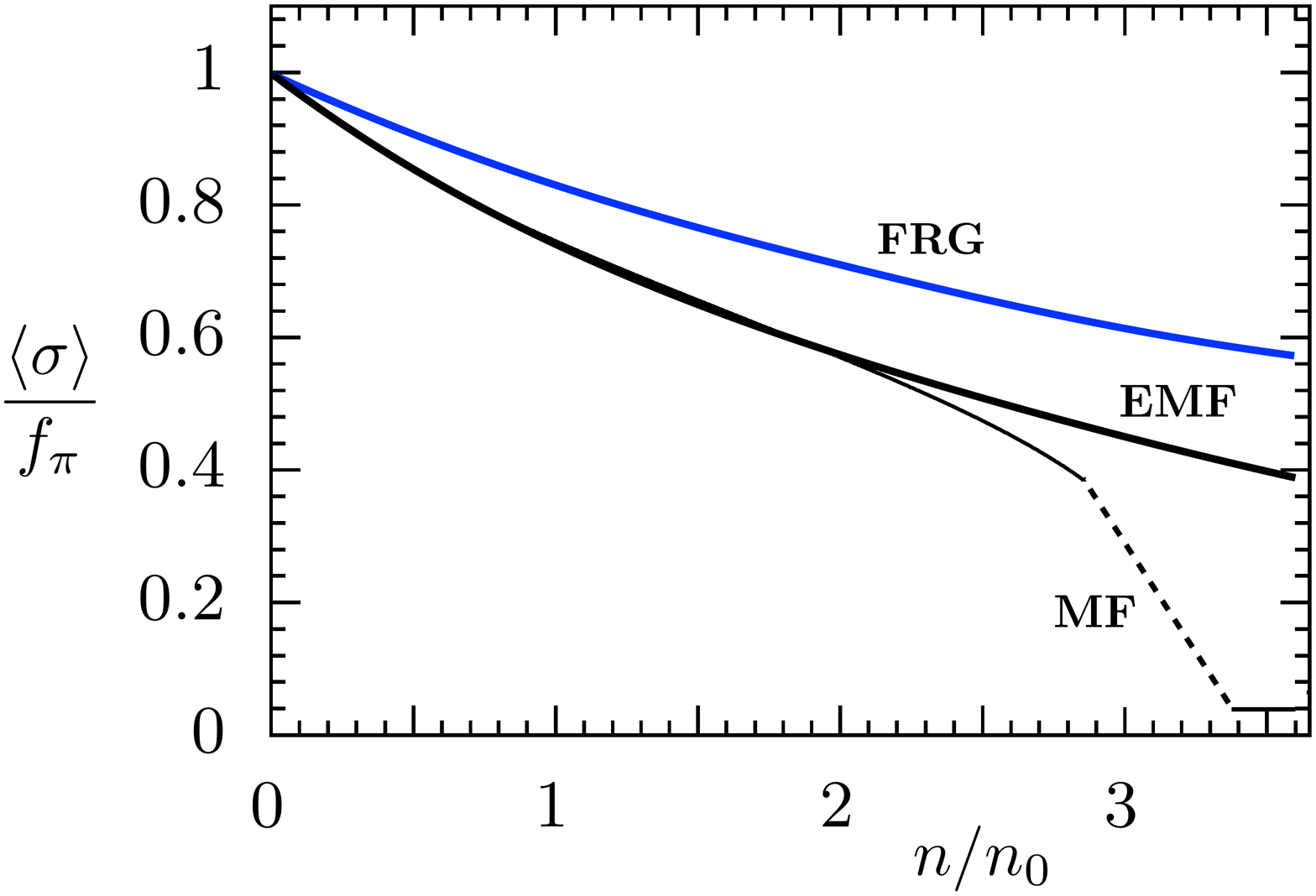}
\caption{Chiral order parameters for pure neutron matter at temperature $T=0$ as a function of neutron density. Legends of the curves are the same as in Fig.\,\ref{fig:3}.}
\label{fig:7}
\end{center}
\end{figure*}

The stabilisation of the chiral order parameter by fluctuations is further enhanced in the full FRG scenario \cite{Drews2015,Drews2017} as indicated by the corresponding curves in Figs.\,\ref{fig:3} and \ref{fig:4}. With the input parameters  (\ref{eq:FRGparameters}) reproducing nuclear matter ground state properties, the pattern of the liquid-gas phase transition from symmetric to neutron-rich matter is very close to the one in the EMF scheme, Fig.\,\ref{fig:2}. Marginal differences occur in the critical temperature ($T_{\textrm{crit}} = 18.3$ MeV (FRG) vs. $T_{\textrm{crit}} = 17.5$ MeV (EMF)), and the disappearance of the liquid-gas coexistence region at a proton fraction $x \simeq 0.04$ (FRG) instead of $x \simeq 0.02$ (EMF). The exact values depend on the selected parametrisation which includes a certain amount of freedom. 

As already pointed out the FRG framework is richer in dynamical content than EMF. Beyond nucleonic zero-point energies it includes loop effects from pions, sigma bosons and nucleons on the chiral potential ${\cal U}_{k=0}^{(0)}$. These mechanisms shift the chiral transition to even higher densities. 

The high-density behaviour of $\langle\sigma\rangle$, shown for the EMF and FRG scenarios in Fig.\,\ref{fig:5}, suggests a smooth chiral crossover around $n\sim 6\,n_0$ for EMF and at even much higher densities for FRG.
Of course, at such high densities nucleons supposedly overlap and release their quark contents. Also, the ChNM model was adjusted to reproduce properties of the liquid-gas phase transition and the potential was expanded around $\chi_0 = 1/2\,f_{\pi}^2\,$. Hence, if $\langle\sigma\rangle$ becomes too small the model reaches its limit of applicability.  
However, the qualitative feature of a chiral crossover induced by fluctuations, instead of a first-order chiral phase transition, is expected to persist. 

\begin{figure*}[t]
\begin{center}
\includegraphics[height=60mm,angle=-00]{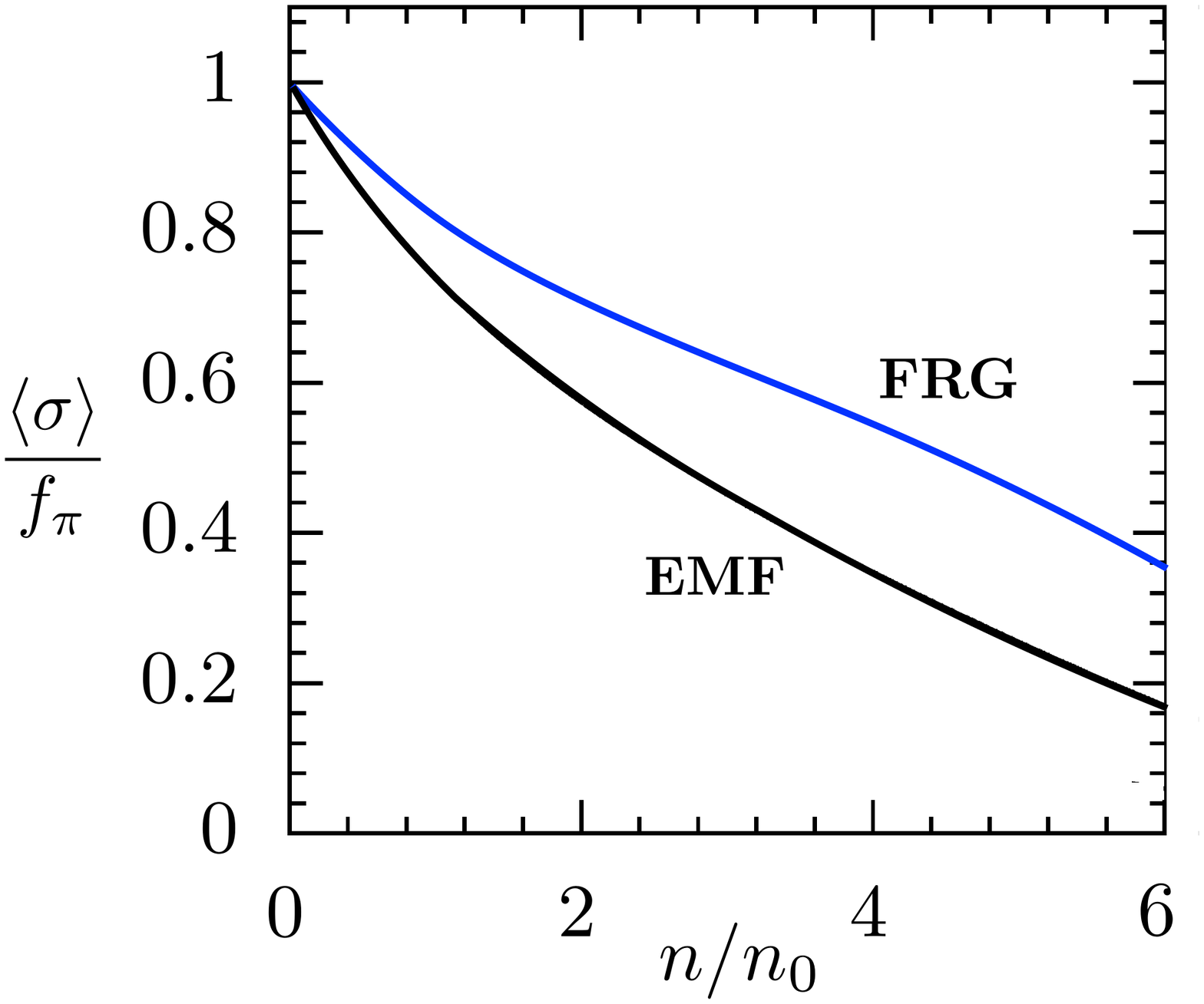}
\caption{Chiral order parameters in neutron matter using the ChNM model as described in the text. Shown in particular is the behaviour at high neutron density (in units of $n_0 = 0.16$ fm$^{-3}$) for the EMF and FRG scenarios as indicated.}
\label{fig:8}
\end{center}
\end{figure*}

\begin{figure*}[t]
\begin{center}
\includegraphics[height=60mm,angle=-00]{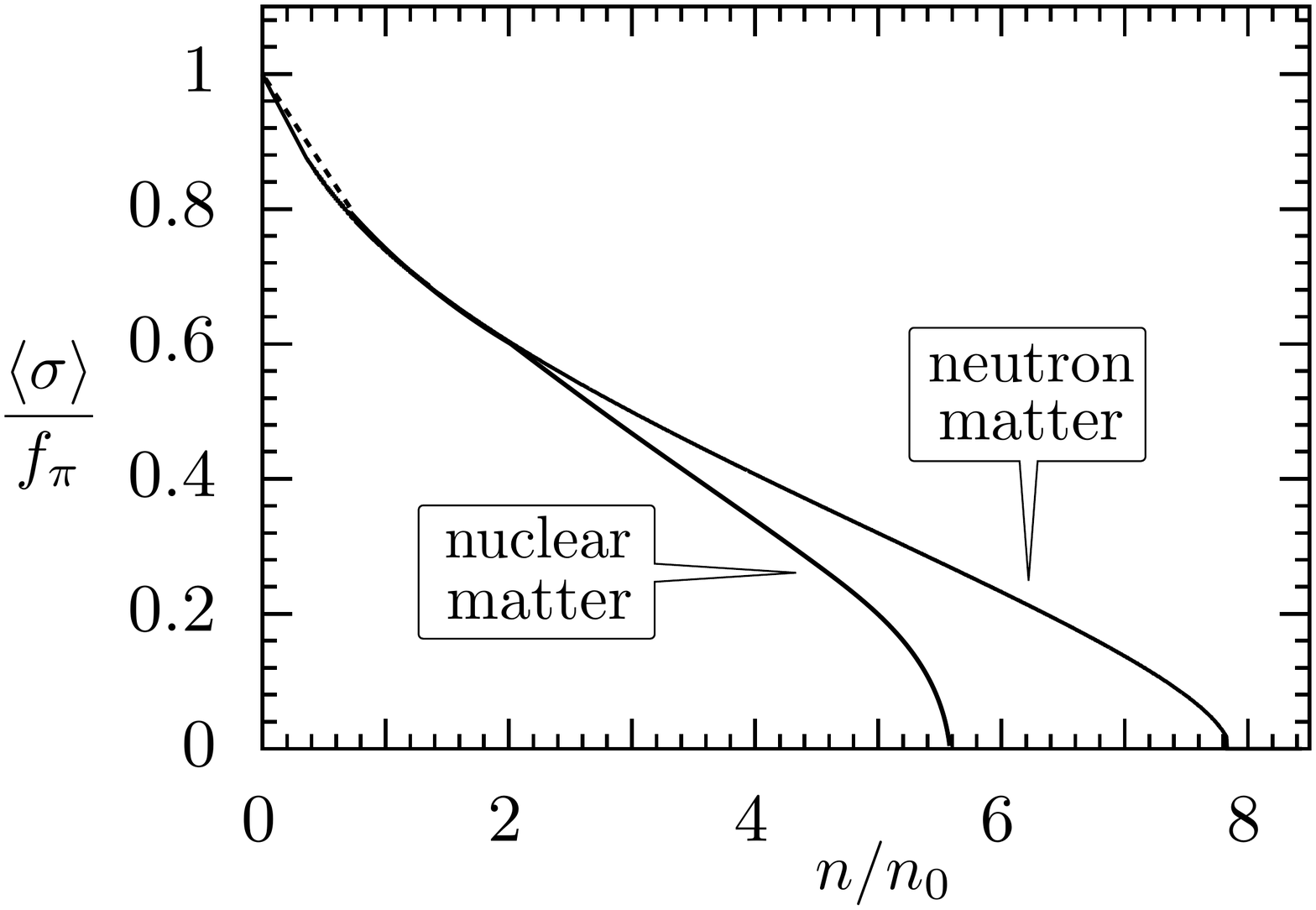}
\caption{Chiral order parameters in the chiral limit ($m_\pi\rightarrow 0$) for symmetric nuclear matter and neutron matter, respectively, at $T=0$ as functions of baryon density. The curves showing 2nd order chiral restoration phase transitions at high densities have been calculated using the ChNM model in the EMF approximation (mean-field approximation plus vacuum loop corrections) as described in the text.}
\label{fig:9}
\end{center}
\end{figure*}

\subsection{Neutron matter}

As a prerequisite before entering the discussion of chiral phases in neutron matter, Fig.\,\ref{fig:6} shows the energy per particle at low density calculated using the EMF and FRG schemes, in comparison with results of calculations based on a chiral $N^3LO$ nucleon-nucleon interaction with inclusion of three- and four-body contributions \cite{Drischler2019} (see also \cite{Holt2017}). The ChNM model combined with FRG closely resembles state-of-the-art results of $N^3LO$ chiral effective field theory at low densities within uncertainties, so that one can proceed to higher density with some confidence.

Apart from the liquid-gas transition, the chiral order parameter in neutron matter shows a qualitatively similar behaviour as in symmetric nuclear matter. In the MF limit there would be a first-order chiral phase transition starting from a density slightly below $3\,n_0$. As a function of neutron chemical potential this first-order transition occurs at $\mu_n \simeq 1.2$ GeV. Adding vacuum fluctuations in the EMF scheme stabilizes the system and induces a smooth behaviour of $\langle\sigma\rangle$. The full FRG calculation with its repulsive loop corrections provides further stabilization and moves the transition to chiral symmetry restoration in the form of a crossover to densities way beyond $6\,n_0$, as illustrated in Figs.\,\ref{fig:7} and \ref{fig:8}. 

At this point one should recall that central densities in heavy $(\sim 2\,M_\odot)$ neutron stars are expected to be typically in a range $5-6\,n_0$ if their radii exceed 10 km \cite{Hell2014,Jiang2019}. A first-order chiral phase transition occurring below that density range would cause the EoS to be too soft for generating the necessary high pressure to keep such heavy stars stable. The FRG version of an EoS of neutron star matter (including beta equilibrium) based on the ChNM model acquires the necessary stiffness to support $2\,M_\odot$ stars with radii around $R \simeq 12$ km \cite{Drews2017,FW2019}. The present EMF version using the input in Eq.\,(\ref{eq:EMFparameters}) barely misses this high-density constraint, the maximum neutron star mass generated by the corresponding EoS being $1.84\,M_\odot$. We have checked that a minor modification of the effective potential, raising the maximum power of the polynomial expansion (\ref{eq:potential}) to $N=6$ with ``natural" choices of small parameters $a_5, a_6$, can easily improve the high-density EoS to achieve the required stiffness, but leads to a worse description of the nuclear phenomenology. However, this is not in the main focus of the present work.

\begin{figure*}[t]
\begin{center}
\includegraphics[height=60mm,angle=-00]{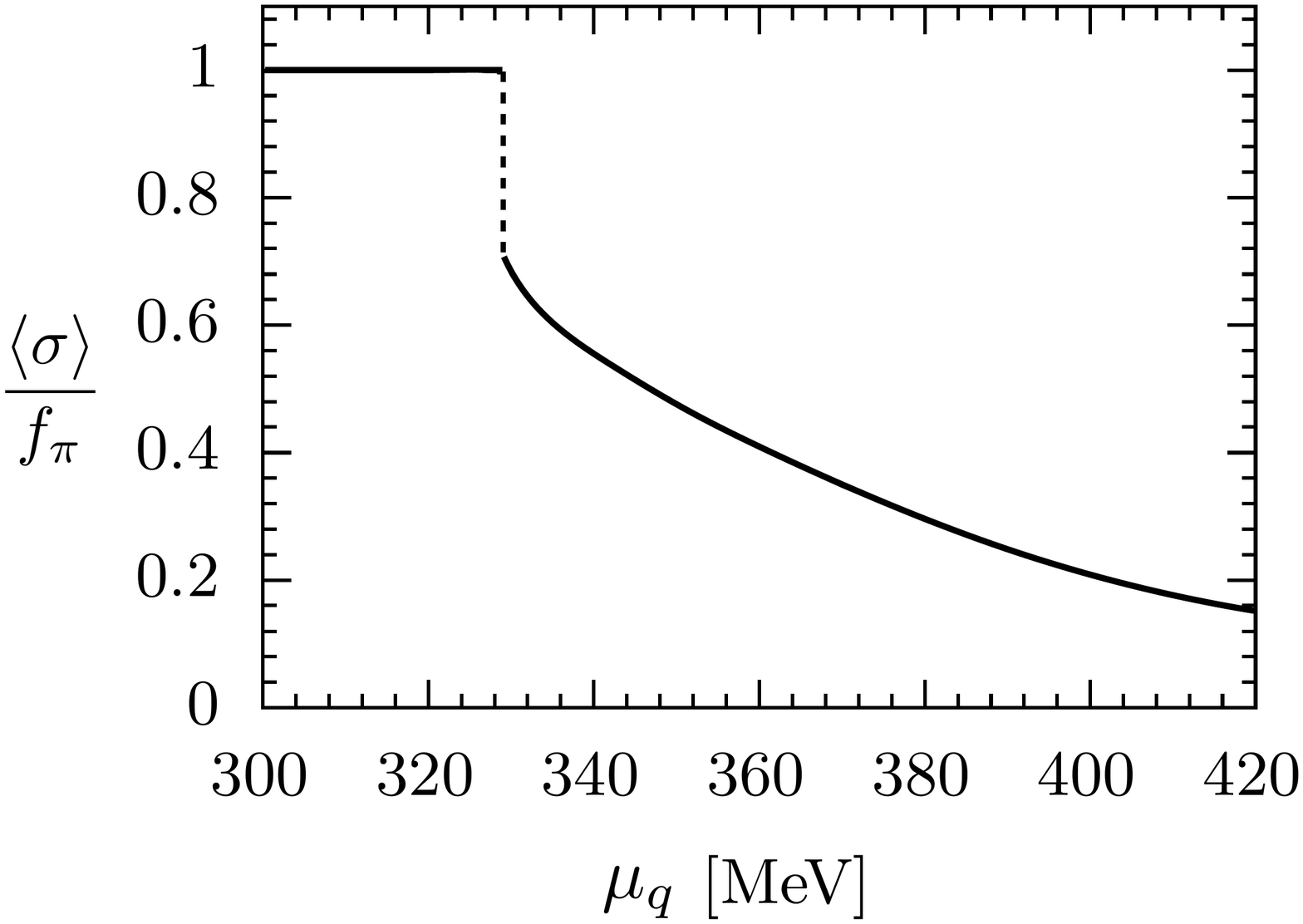}
\caption{Chiral order parameter for symmetric quark matter at temperature $T=5$ MeV as function of quark chemical potential. Result of an FRG calculation using a chiral quark-meson model with input as in Ref.\,\cite{Pereira2020} and zero vector coupling. }
\label{fig:10}
\end{center}
\end{figure*}
\subsection{Chiral limit}

It is instructive to examine the previous scenarios in the chiral limit, i.e. for zero pion mass and in the absence of the term in the action that describes explicit chiral symmetry breaking. In this case the pion decay constant in vacuum is reduced to $f_\pi = 86$ MeV and the nucleon mass is lowered correspondingly. A reparametrisation is performed so that selected nuclear constraints (such as $E_0/A = -16$ MeV) are still fulfilled in order to have a common baseline. The saturation density for symmetric nuclear matter shifts to $0.7\,n_0$, assuming that the input sigma mass stays unchanged. For symmetric nuclear matter the critical parameters of the liquid-gas phase transition change moderately: the critical temperature increases to 18.3 MeV. Neutron matter now also becomes weakly bound and develops a weak first-order liquid-gas phase transition with low critical temperature at a neutron chemical potential $\mu_n \simeq 880 - 890$ MeV. In the EMF scheme that we use here for demonstration, the chiral transition appears as a second-order phase transition at very high critical densities (at 5.6  $n_0$ in symmetric nuclear matter and at 7.8  $n_0$ in neutron matter). This is displayed in Fig.\,\ref{fig:9}.

\section{Comment on a chiral quark-meson model}

Much previous work in quest for a first-order chiral phase transition was performed using various versions of chiral quark-meson (ChQM) models. In these models the isospin $SU(2)$ doublet quark field, $\psi = (u,d)$, replaces the nucleon field $\Psi$ of previous Sections. A recent example of such a ChQM model combined with FRG methods has been employed in Ref.\,\cite{Pereira2020}. Their study is concerned with the role of vector interactions on the thermodynamics of a first-order phase transition which they interpret as a chiral phase transition. With the approximations introduced, such as the treatment of vector fields in mean-field approximation, the model is analogous to the ChNM $k$-dependent action (\ref{eq:k-action}):
\begin{eqnarray}
\Gamma_k &=&\int^{1/T}_0 dx_4\int{d^3x} \Bigl\{\bar{\psi}\left[\gamma_\mu\partial_\mu + g_s(\sigma + i\gamma_5\,\boldsymbol{\tau\cdot\pi})\right]\psi \nonumber \\
&+&\psi^\dagger\left(\boldsymbol{\mu} - g_v\,v - g_w\,\tau_3\,w \right)\psi + {\frac12}\left(\partial_\mu \sigma \partial_\mu \sigma + \partial_\mu \boldsymbol{\pi}\cdot\partial_\mu \boldsymbol{\pi}\right) \nonumber\\
&+& {\cal U}_k(T,\mu_u,\mu_d;\sigma, \boldsymbol{\pi}, v,w)\Bigr\} ~,
\label{eq:k-actionQM}
\end{eqnarray}
with
\begin{eqnarray}
\boldsymbol{\mu} = \begin{pmatrix}
\mu_u & \\
& \mu_d
\end{pmatrix}~,
\end{eqnarray} 
and
\begin{equation}
{\cal U}_k = {\cal U}_k^{(0)}(\chi) -c\,\sigma - {1\over 2}m_v^2(v^2+w^2)~.
\end{equation}
At a UV scale $k_{UV} = \Lambda = 1$ GeV the chiral invariant part of the potential is parametrised in the simplified form
\begin{equation}
{\cal U}^{(0)}_{k=\Lambda} =  m_\Lambda^2\,\chi + \lambda\,\chi^2~.
\end{equation}
The symmetry breaking term with $c = m_\pi^2 f_\pi$ differs from the one used previously in ChNM model  just by an irrelevant additive constant. Effective chemical potentials for $u$ and $d$ quarks, introduced as
\begin{eqnarray}
\bar{\mu}_u(\chi) &=& \mu_u - g_v\,\bar{v}_k(\chi) - g_w\,\bar{w}_k(\chi)~,\nonumber\\
\bar{\mu}_d(\chi) &=& \mu_d - g_v\,\bar{v}_k(\chi) + g_w\,\bar{w}_k(\chi)~,
\end{eqnarray}
depend on the FRG running scale $k$ and the fields $\bar{v}_k$ and $\bar{w}_k$ refer to those that minimize the action $\Gamma_k$ at each $k$. Dynamical quark masses $m_q$ with $q=u,d$ are generated as
\begin{equation}
m_q = g_s\langle\sigma\rangle~.
\end{equation}
Solving FRG flow equations and extracting the effective action in the IR limit $k=0$, the choice of initial parameters in Ref.\,\cite{Tripolt2018,Pereira2020}, $g_s = 4.2,\,m_\Lambda = 0.97$ GeV, $\lambda = 10^{-3}$,
produces a dynamical (constituent) quark mass $m_q = 388$ MeV, a sigma mass $m_\sigma = 607$ MeV together with physical values for the pion mass and decay constant in vacuum \cite{footnote5}.

In this ChQM model the authors of Ref.\,\cite{Pereira2020} performed systematic studies of the appearance and properties of a first-order phase transition and its dependence on the vector field couplings in various circumstances. They referred to this as a chiral phase transition. Given the experience with the role of fluctuations in FRG-based calculations as they tend to inhibit a first-order chiral phase transition, we wish to examine this claimed interpretation in more detail. For this purpose we have repeated their calculation, reproducing their result in symmetric quark matter with equal number of $u$ and $d$ quarks, where a first-order phase transition indeed appears, e.g. at $T=5$ MeV and a quark chemical potential $\mu_q = 328$ MeV for vanishing vector coupling, $g_v = 0$. Here the vector coupling strengths are free, unconstrained parameters, unlike the situation in the ChNM model where they are constrained by the requirement to reproduce nuclear phenomenology. For an isoscalar vector coupling $g_v = 3.1$ (corresponding to $G_v = g_v^2/m_\omega^2 = 0.62$ fm$^2$ if the isoscalar vector boson mass is identified with the physical mass of the $\omega$ meson), this transition shifts to $\mu_q = 345$ MeV at $T = 5$ MeV. These critical quark chemical potentials are systematically lower than the dynamical quark mass in vacuum, indicating a binding situation with two degenerate minima in the effective potential ${\cal U}_{k=0}(\sigma)$. 

So far so good - but is this indeed a {\it chiral} phase transition as mentioned in Ref.\,\cite{Pereira2020}? To examine this issue the chiral order parameter $\langle\sigma\rangle$ needs to be investigated. A representative example is shown in Fig.\,\ref{fig:10}, for symmetric quark matter at $T= 5$ MeV and the case with no vector coupling, $g_v = 0$. The existence of a first-order phase transition is evident, but the chiral order parameter does not vanish: chiral symmetry is spontaneously broken throughout the $\mu_q$ interval on display which reaches up to high baryon densities (note that the baryon chemical potential is $\mu = 3\,\mu_q$). Instead, the figure shows a characteristic pattern similar to that of the liquid-gas phase transition in nuclear matter. This suggests an interpretation in terms of a liquid-gas transition in matter formed by constituent quarks as fermionic quasiparticles, but at relatively low baryon densities where quarks are not yet expected to be active degrees of freedom. This interpretation is in accordance with previous analyses \cite{Schaefer2005,Tripolt2018} and qualitatively similar features are seen in various other situations with non-vanishing vector couplings. 

However, if the chiral quark-meson model is treated in mean-field (MF) approximation, the chiral order parameter signals indeed a first-order phase transition \cite{Scavenius2001b,Schaefer2007b,Nickel2009}. After including a logarithmic fermionic vacuum term in EMF, which takes quark fluctuations into account, the expectation value $\langle\sigma\rangle$ is stabilized at larger quark chemical potentials \cite{Zacchi2018}. A similar behaviour is found for Polyakov loop extended quark-meson models \cite{Chatterjee2012,Gupta2012}. By analogy, this corresponds very well to our observations using the chiral nucleon-meson model.

In essence, a first-order {\it chiral} phase transition that occurs in several ChQM models in mean-field calculations is again inhibited and shifted to high densities, possibly converted to a smooth crossover, by the fluctuations as they are treated explicitly in the FRG approach. 

\section{Summary and conclusions}

The present investigation has addressed the following question: how does a possible first-order chiral phase transition in dense baryonic matter at zero temperature react to fluctuations incorporated in an effective potential? As a prototype framework to deal with this issue we have employed a chiral $SU(2)_L\times 
SU(2)_R$ nucleon-meson model. Short-distance dynamics of the nucleon-nucleon interaction are described by additional vector boson couplings. With a limited set of parameters this model is capable of reproducing empirical nuclear bulk properties and the thermodynamics of the nuclear liquid-gas phase transition. This consistency with nuclear phenomenology is a necessary requirement for meaningful extrapolations to higher baryon densities, beyond the density of nuclear matter in its ground state, $n_0 = 0.16$ fm$^{-3}$.

The central quantity of interest is then the chiral order parameter, in our case the expectation value $\langle\sigma\rangle$ of the scalar field accompanying the pion (the chiral Nambu-Goldstone boson). Its normalisation is $\langle\sigma\rangle_{vac} = f_\pi \simeq 93$ MeV, the pion decay constant associated with the axial current transition matrix element from the vacuum to a one-pion state. We are thus interested in the behaviour of the pion decay constant in a baryonic medium as function of density and temperature. In the present work we focus primarily on the case of cold $(T=0)$ and highly compressed matter composed of nucleons. A vanishing of $\langle\sigma\rangle$ at high density implies restoration of chiral symmetry in its Wigner-Weyl realisation, along with a vanishing nucleon mass that is rigidly coupled to the scalar field, $M = g\langle\sigma\rangle$. 

The results can be summarised as follows:
\begin{itemize}
\item{{\bf i)} In the simplest mean-field (MF) approximation, both symmetric nuclear matter and pure neutron matter feature chiral first-order transitions at relatively low baryon chemical potentials or densities. With given input parameters reproducing empirical nuclear properties, a chiral first-order transition in nuclear matter would occur at a chemical potential $\mu_c\simeq 945$ MeV, corresponding to a coexistence range of baryon densities, $1.5\,n_0\lesssim n \lesssim 3.1\,n_0$. For neutron matter the corresponding transition would occur at a neutron chemical potential $\mu_{n,c} \simeq 1.2$ GeV, corresponding to a coexistence range of neutron densities, $2.8\,n_0\lesssim n_n \lesssim 3.3\,n_0$. At such low densities the existence of a first-order chiral phase transition would imply a dense matter equation-of-state that would be too soft to support the heaviest observed neutron stars.}\\

\item{{\bf ii)} As a next step, vacuum fluctuations in the form of a renormalised nucleonic zero-point energy density have been included in an extended mean-field (EMF) approach. This minimal inclusion of fluctuations beyond MF takes into account the one-loop fermionic effective potential. It already induces the remarkable effect of shifting the chiral transition to much higher densities and converting it into a smooth crossover. In nuclear matter, chiral symmetry now remains in its spontaneously broken Nambu-Goldstone realisation up to densities beyond $5\,n_0$. For neutron matter this spontaneously broken phase persists up to neutron densities well above $6\,n_0$. An interesting special case is the chiral limit, taking the pion mass to zero and reducing at the same time the pion decay constant in vacuum to $f_\pi^{(0)} \simeq 86$ MeV. In this limit a second order chiral phase transition occurs. With vacuum fluctuations included in the EMF scheme, the corresponding critical densities are located at $5.6\,n_0$ and $6.8\,n_0$ in nuclear and neutron matter, respectively. Of course, all these statements should be taken as pointing out a qualitative trend rather than a quantitative measure, given that nucleons are expected to start percolating and releasing their quark contents over wider domains at such high densities.}\\

\item{{\bf iii)} Within the same chiral model, a further comparison with calculations using non-perturbative functional renormalisation group (FRG) methods points to an even stronger importance of fluctuations in stabilizing the tendency towards chiral restoration at high baryon densities. These results include not only fermionic vacuum fluctuations but also loop and thermal corrections involving pions, sigma bosons and nucleons. In both nuclear and neutron matter, the chiral order parameter now stays significantly further away from zero, at about $40\%$ of its vacuum value at densities as high as $n \sim 6\,n_0$.}\\

\end{itemize}

This situation is qualitatively reminiscent of early studies that postulated an abnormal Lee-Wick phase  \cite{LeeWick1974} in dense matter, signalled by a vanishing dynamical nucleon mass controlled by the expectation value of a scalar field. Later one-loop corrections were added and the role of many-body forces was studied, with the conclusion that in the presence of such fluctuations beyond mean-field, the Lee-Wick phase sets in only at very high densities \cite{Nyman1976}. 

In summary, we have pointed out that the discussion of a possible first-order chiral phase transition in the equation-of-state of dense baryonic matter requires a systematic treatment of fluctuations beyond mean-field approximation. These fluctuations reflect repulsive loop effects which grow with increasing density and tend to stabilise the trend towards a transition to chiral symmetry restoration. We have conducted these studies for the example of a chiral nucleon-meson model, but our short comment regarding the chiral quark-meson model indicates that these conclusions may be of a more general nature.


\begin{thebibliography}{99}
\bibitem{footnote0}
The ChNM Lagrangian is written in terms of the nucleon field $\Psi$. For the ChQM model to be commented on at a later stage, simply replace $\Psi$ by the quark doublet field, $\psi= (u,d)$. The Euclidean Dirac matrices satisfy the anticommutation relation $\{\gamma_\mu,\gamma_\nu\} = 2\delta_{\mu\nu}$ with $\mu, \nu = 1,\,...\,,4$\,.
\bibitem{footnote3}
In practice we use $k_{UV} = 1.4$ GeV which extends the covered momentum region slightly above the chiral scale $\Lambda_\chi$.
\bibitem{footnote5}
Note that, in contrast to the ChNM model, the parameters of the ChQM model can not be fixed to low-density phenomenology except for $m_\pi$ and $f_\pi$.
\end{thebibliography}
\end{document}